\documentclass[preprint2]{aastex}

\usepackage{natbib}
\usepackage{amssymb,amsmath}
\textheight=\baselinestretch\textheight
\ifdim\textheight>25.2cm\textheight=25.0cm\fi

\topskip\baselineskip
\maxdepth\baselineskip
\let\tighten=\relax
\let\tightenlines=\tighten

\newcommand{\cmmthree}{\mbox{cm$^{-3}$}}

\newcommand{\kms}{\mbox{km\,s$^{-1}$}}
\newcommand{\degree}{\mbox{$^{\circ}$}}
\newcommand{\degrees}{\mbox{$^{\circ}$}}
\newcommand{\msun}{\mbox{M$_{\odot}$}}

\newcommand{\microns}{\mbox{$\mu$m}} 
\newcommand{\jybm}{\mbox{Jy\,beam$^{-1}$}}
\newcommand{\rmtwo}{\mbox{rad\,m$^{-2}$}}
\newcommand{\muG}{\mbox{$\mu$G}}

\newcommand{\mjbm}{\mbox{mJy\,beam$^{-1}$}}
\newcommand{\pccmmsix}{\mbox{pc\,cm$^{-6}$}}
\newcommand{\pccmmthree}{\mbox{pc\,cm$^{-3}$}}

\newcommand{\halpha}{\mbox{H\,$\alpha$}}

\newcommand{\hii}{\mbox{H\,{\scriptsize II}}}
\newcommand{\hi}{\mbox{H\,{\scriptsize I}}}

\slugcomment{Polarisation in the Gum Nebula}

\shorttitle{Polarisation in the Gum Nebula}
\shortauthors{C.~R.~Purcell et al.}

\begin{document}

\title{A radio-polarisation and rotation measure study of the Gum Nebula and its environment}

\author{
C.~R.\,Purcell\altaffilmark{1}$^{\dagger}$,
B.~M.~Gaensler\altaffilmark{1},
X.~H.~Sun\altaffilmark{1},
E.~Carretti\altaffilmark{2}, 
G.~Bernardi\altaffilmark{3,\,4},
M.~Haverkorn\altaffilmark{5}
M.~J.~Kesteven\altaffilmark{2},
S.~Poppi\altaffilmark{6},
D.~H.~F.~M.~Schnitzeler\altaffilmark{7} \and
L.~Staveley-Smith\altaffilmark{8,\,9} }
\altaffiltext{$\dagger$}{\bf cormac.purcell@sydney.edu.au}
\altaffiltext{1}{Sydney Institute for Astronomy (SIfA), School of
Physics, The University of Sydney, NSW 2006, Australia}
\altaffiltext{2}{ATNF, CSIRO Astronomy and Space Science, PO Box 76, Epping,
  NSW 1710, Australia}
\altaffiltext{3}{SKA SA, 3rd Floor, The Park, Park Road, Pinelands,
  7405, South Africa}
\altaffiltext{4}{Department of Physics and Electronics, Rhodes
  University, PO Box 94, Grahamstown, 6140, South Africa}
\altaffiltext{5}{Department of Astrophysics/IMAPP, Radboud University
  Nijmegen, PO Box 9010, NL-6500 GL Nijmegen, the Netherlands}
\altaffiltext{6}{INAF – Osservatorio Astronomico di Cagliari, St. 54
  Loc. Poggio dei Pini, I-09012 Capoterra (CA), Italy}
\altaffiltext{7}{Max-Planck-Institut f{\"u}r Radioastronomie, Auf dem
H{\"u}gel 69, D-53121 Bonn, Germany}
\altaffiltext{8}{International Centre for Radio Astronomy Research,
   M468, University of Western Australia, 35 Stirling Highway,
   Crawley, Western Australia 6009, Australia}
\altaffiltext{9}{ARC Centre of Excellence for All-sky Astrophysics
  (CAASTRO), M468, University of Western Australia, 35 Stirling
  Highway, Crawley, Western Australia 6009, Australia} 

\begin{abstract}
The Gum Nebula is $36\degrees$-wide shell-like emission nebula at a
distance of only $\sim450$\,pc. It has been hypothesised to be
an old supernova remnant, fossil $\hii$~region, wind-blown bubble, or
combination of multiple objects. Here we investigate the
magneto-ionic properties of the nebula using data from recent
surveys: radio-continuum data from the NRAO VLA and S-band
Parkes All Sky Surveys, and $\halpha$ data from the Southern H-Alpha
Sky Survey Atlas. We model the upper part of the nebula as a spherical
shell of ionised gas expanding into the ambient medium. We perform a
maximum-likelihood Markov chain Monte-Carlo fit to the NVSS rotation
measure data, using the $\halpha$ data to constrain average electron
density in the shell $n_e$. Assuming a latitudinal background gradient
in RM we find $n_e=1.3^{+0.4}_{-0.4}\,\cmmthree$, angular radius
$\phi_{\rm outer}=22.7^{+0.1}_{-0.1}\,{\rm degrees}$, shell thickness
$dr=18.5^{+1.5}_{-1.4}\,{\rm pc}$, ambient magnetic field strength
$B_0=3.9^{+4.9}_{-2.2}\,\muG$ and warm gas filling factor
$f=0.3^{+0.3}_{-0.1}$. We constrain the local, small-scale
($\sim260$\,pc) pitch-angle of the ordered Galactic magnetic field to
$+7\degrees\lesssim\wp\lesssim+44\degrees$, which represents a 
significant deviation from the median field orientation on kiloparsec
scales ($\sim -7.2\degrees$). The moderate compression factor
$X=6.0\,^{+5.1}_{-2.5}$ at the edge of the $\halpha$ shell implies
that the `old supernova remnant' origin is unlikely. Our results
support a model of the nebula as a $\hii$~region around a wind-blown
bubble. Analysis of depolarisation in 2.3\,GHz S-PASS data is
consistent with this hypothesis and our best-fitting values agree well
with previous studies of interstellar bubbles.
\end{abstract}



\section{Introduction}\label{sec:intro}
Observations of atomic $\hi$, molecular clouds  and photo-dissociation
regions  in the Galaxy have shown that gas in a
wide range of environments is gathered into spheres, bubbles or shell-like
structures (e.g., \citealt{Jackson2006}; \citealt{Churchwell2006};
\citealt{McClure-Griffiths2009}). Most of these objects are formed 
by physical processes associated with the evolution of high-mass stars
($>8\,\msun$). During their time on the main-sequence, such stars emit
high fluxes of ultra-violet photons and fast winds of particles that
ionise expanding \hii~regions, evacuate low-density cavities and sweep
gas into shells in a `snow-plough' effect. At the end of their lives
the stars eject their outer layers, before exploding as supernovae,
driving strong shocks into the interstellar medium (ISM). OB-type
stars generally form in clusters,  
so the combined action of stellar winds and coeval supernova
explosions can give rise to `supershells', hundreds of parsecs in size
(e.g., \citealt{Moss2012}). Supernovae and supershells are thought to
power the circulation of material into the Galactic halo
(\citealt{Dove2000}; \citealt{Reynolds2001},
\citealt{Pidopryhora2007}) and play a leading role in sculpting the
fractal structure of gas in the Galactic disk. With energies greater
than $\sim10^{51}$\,ergs, supernovae are also believed to be the main
driver of turbulence in the disk \citep{McCray1979} and have
been shown to trigger new episodes of star-formation when shocks
overrun and compress pre-existing clumps of molecular gas
\citep{Reipurth1983, Oey2005}.

Spheres or bubbles of plasma also present an excellent opportunity to
probe conditions in the ISM. Studies of individual objects can yield
information on the conditions in the medium into which they are
expanding and on their progenitors. Of particular interest are recent
works that have used observations of Faraday rotation to derive
the magneto-ionic properties of bubbles and their local ISM
(e.g.  \citealt{Kothes2009}; \citealt{Whiting2009};
\citealt{Harvey-Smith2011}; \citealt{Savage2013}). As the bubbles
expand they interact with the ordered magnetic field of the Galaxy, 
compressing the ambient medium and the field parallel to the shock
front. The resulting field geometry is a function of the pre-existing
field configuration and the rate of expansion, leading to a unique
rotation measure (RM) signature on the sky. While observations of RMs
from extra-galactic point sources yield only the average line-of-sight
field strength, modelling the RM-signature of a supernova 
remnant or $\hii$~region is one of the few ways to measure the local
magnetic field on scales of a few hundred parsecs. Such measurements
are essential anchor-points for studies of the large-scale Galactic
field (e.g., \citealt{Kothes2009}).

If the line-of-sight magnetic field strength is known, Faraday
rotation is a good tool to measure density jumps in the ionised ISM as
$n_e\propto{\rm RM}$.  The density jump present at the 
shell boundary is a key indicator of the type of object powering the
expansion. For example, mass and momentum conservation in the
radiatively cooling shock-fronts of old supernova remnants (SNR)
are expected to lead to very high density jumps at their boundaries
\citep{Shull1987}. In contrast, the ionisation
front of an evolved $\hii$~region created by a cluster of
B-type stars would expand at the local sound speed (typically
$\sim10$\,\kms), creating only a slight density jump. The level of
compression and magnetic field strength in the ionised gas have
profound implications for whether star formation is triggered or
suppressed by the passing shock.

In this paper we present a study of one of the most prominent bubbles
in the southern sky: the Gum Nebula. We start in
$\S\ref{sec:gum_history}$ by reviewing the literature on the nebula,
summarising its properties and theories of origin. In
$\S\ref{sec:datasets}$ we introduce the datasets and images
used in this work. We go on to describe our ionised shell model and
analysis techniques in $\S\ref{sec:analysis}$. We present the results 
of fitting the model to the rotation measure data in
$\S\ref{sec:results}$, where we derive strength and direction of the
magnetic field, and the density jump across the edge of the
nebula. Discussion and further analysis of depolarisation at radio
wavelengths are presented in $\S\ref{sec:discussion}$. Finally, we
present our conclusions in $\S\ref{sec:summary}$ and suggest
future avenues of investigation.

\subsection{The Gum Nebula}\label{sec:gum_history}
\begin{figure*}
  \centering
  \includegraphics[width=18.0cm, trim=0 0 0 0]{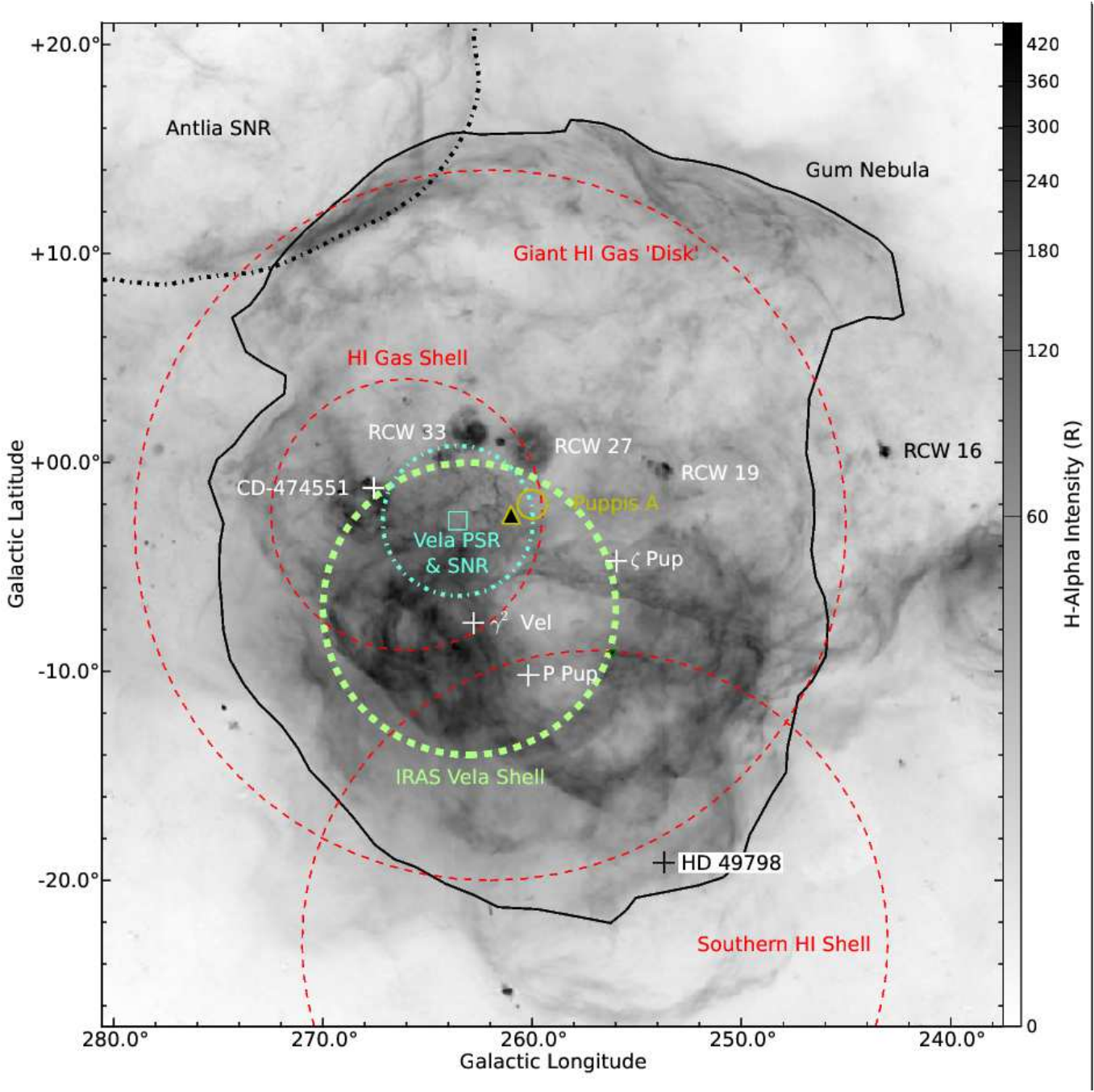}
  \caption{\small An $\halpha$ image of the Gum Nebula
    \citep{Finkbeiner2003} annotated with significant objects identified
    in the literature. OB-type stars are marked with `+' symbols,
    the Vela pulsar with a square and the kinematic centre of the nebula
    derived by \citet{Woermann2001} with a triangle. Boundaries of
    gas-disks or shells are marked using coloured circles. See
    $\S\ref{sec:gum_history}$ for details. References for the
    annotated objects are as follows:  Antlia SNR:
    \citealt{McCullough2002}; Giant $\hi$ Gas Disk:
    \citealt{Reynoso1997}; $\hi$ Gas Shell: \citealt{Dubner1992}; IRAS
    Vela Shell: \citealt{Sahu1993}; RCW-$\hii$~regions:
    \citealt{Rodgers1960}; Vela Pulsar: \citealt{Radhakrishnan1969};
    CD-474551: \citealt{Reed2003}; HD\,49798 and~$\zeta$\,Puppis:
    \citealt{vanLeeuwen2007}; $\gamma^2$\,Velorum:
    \citealt{DeMarco1999}.}
  \label{fig:gum_summary_halpha}
\end{figure*}
The Gum Nebula is one of the largest optical emission nebulae in the
southern sky \citep{Gum1952}. It has an approximately circular
morphology with an angular diameter of $\sim36\degree$
\citep{Chanot1983}, its centre is 
thought to lie at a distance of $\sim$500\,pc from the Sun
and its radius is $\sim130$\,pc \citep{Woermann2001}. Originally
discovered in large-area photographic plates by \citet{Gum1952}, it
dominates modern $\halpha$ maps of the southern Galactic plane (e.g.,
\citealt{Dennison1998}; \citealt{Gaustad2001}; \citealt{Haffner2003}).

\subsubsection{The environment of the Gum Nebula}
Considerable controversy exists in the literature on the origin and
evolution of the Gum Nebula. In part, this is because the nebula
straddles the mid-plane of the Galaxy and its footprint encompasses a
large number of overlapping objects: $\hii$ regions, supernova
remnants (SNRs), OB-associations and molecular clouds. Early
investigations of the nebula (e.g., \citealt{Gum1956};
\citealt{Brandt1971}; \citealt{Alexander1971}; \citealt{Beuermann1973};
\citealt{Reynolds1976b}; \citealt{Weaver1977}; \citealt{Vallee1983})
were limited by the paucity of observations covering the entire
region; however, more recent work has begun to form a clear picture
(e.g., \citealt{Sahu1992}; \citealt{Sahu1993}; \citealt{Duncan1996};
\citealt{Reynoso1997}; \citealt{Woermann2001};
\citealt{Stil2007}). Fig.~\ref{fig:gum_summary_halpha} presents an  
annotated $\halpha$ image of the nebula in Galactic coordinates
\citep{Finkbeiner2003} illustrating the principal structures 
identified to-date. The upper third of the nebula is relatively free
of confusing sources except at $(l,b)\approx(268\degrees,+13\degrees)$
where the $\halpha$ shell overlaps the Antlia supernova remnant (SNR,
\citealt{McCullough2002,Iacobelli2014}). The lower two-thirds contain
the majority of confusing objects, only some of which are directly
associated with the Gum Nebula.

The energy budget of the nebula is dominated by the output of
early-type stars: $\zeta$\,Puppis; an O4f star, and
$\gamma^2$\,Velorum; a Wolf-Rayet star of type WC\,8 with an O7.5\,I
companion \citep{DeMarco1999}.  
$\gamma^2$\,Velorum is embedded in the Vela OB2 association, which
contains a further 81 B-type stars at a mean distance of
$415\pm10$\,pc \citep{deZeeuw1999}. The combined flux from
$\zeta$\,Pup and $\gamma^2$\,Vel is capable of maintaining the
ionisation state of the Gum Nebula \citep{Weaver1977}, however, Vela
OB2 also appears to be creating a smaller bubble within the
Gum Nebula - the IRAS Vela Shell (IVS). The IVS was
identified by \citet{Sahu1993} as a radius\,$=7.5\degrees$ ring-like
structure in the 100\,$\microns$ IRAS Sky Survey Atlas centred on Vela
OB2. It is associated with a thick shell of  $\hi$ \citep{Dubner1992}
and swept-up molecular gas \citep{Churchwell1996}, and has been
interpreted as a wind-blown bubble driven by Vela\,OB2
\citep{Sahu1993}. Two further OB-stars are 
in the field: the O6 star CD-47\,4551 lies well beyond the Gum
Nebula at a distance of $\sim1300$\,pc, too far to be significantly
interacting with the nebula. The star HD\,49798 is a sub-dwarf O6
binary \citep{Bisscheroux1997} located just outside the nebula
($d=600\pm100$\,pc) and is observed to be emitting a wind that is
distorting the lower shell of the Gum Nebula \citep{Reynoso1997}.

A string of $\hii$~regions (e.g., RCW\,19, RCW\,27 \& RCW\,33;
\citealt{Rodgers1960}) are visible in
Fig.~\ref{fig:gum_summary_halpha} bisecting the nebula along the
Galactic plane. Most of these are associated with the Vela
molecular ridge (VMR), a concentration of molecular clouds beyond the
Gum Nebula at a distance of 1\,-\,2\,kpc \citep{May1988,Murphy1991}.
\citet{Reynoso1997} discovered a massive ($1.4\times10^5\,\msun$)
$\hi$ gas disk corresponding to the optical outline of the Gum Nebula 
and speculate that this may be the signature of the expanding rear
wall of the nebula on the VMR. 

The bulk gas motions and excitation conditions of the Gum Nebula shell
have been measured via spectroscopy of optical emission lines. Spectra of
$\halpha$, [N\,{\scriptsize II}]\,$\lambda6584$, 
[O\,{\scriptsize II}]\,$\lambda5007$ and [He\,{\scriptsize I}]
$\lambda5876$ taken by \citet{Reynolds1976a} suggested that much of
the emitting gas is confined to a shell of radius$\,=\,125$\,pc with an
expansion velocity $\sim20\,$\kms, a thickness
$L\,\approx\,15-30$\,pc and a temperature of 11300\,K. The expansion
velocity was later updated to a value $\sim10\,\kms$
and the excitation conditions in the Gum shell measured to be
consistent with an $\hii$~region (\citealt{Wallerstein1980};
\citealt{Srinivasan1987}).

The kinematics of the Gum Nebula have also been studied via
observations of cometary globules: dense accretions of molecular gas 
and dust \citep{Hawarden1976, Sandqvist1976, 
  Zealey1979, Reipurth1983, Sahu1988, Sahu1992, Sahu1993}. The
most comprehensive analysis of the nebula kinematics was carried
out by \citet{Woermann2001} who found that the best-fitting model of the
neutral gas (including OH masers and molecular clouds) was an asymmetric
expanding shell whose front face is expanding faster than the rear
($14\,\kms$ versus $8.5\,\kms$). The 
runaway O-star $\zeta$\,Puppis was within $<0.5\degree$ of the
expansion centre ($l=261\degrees$, $b=-2.5\degrees$) approximately
$\sim1.5$\,Myr ago, leading to speculation that its companion star
exploded, ejected $\zeta$\,Puppis and created the Gum
Nebula. \citet{Woermann2001} question if the arc of $\halpha$ emission
at $b>10\degree$ is part of the nebula, as it lies offset in Galactic
latitude from the best-fitting neutral shell. However, we note that
the upper part of the nebula is not well sampled by any of the
datasets used. Only one data-point from that study (from a diffuse
molecular cloud) lies at $b>10\degrees$, so fits to the upper nebula
are poorly constrained.

 \citet{Duncan1996} estimated that synchrotron emission is responsible
 for only 10 to 20 percent of the total-power from the nebula in their
 2.4\,GHz single-dish map, which covered the interior region
 ($|b|<5\,\degree$). The hydrogen radio recombination lines
 H156$\alpha$ and H139$\alpha$ were detected by \citet{Woermann2000}
 at four positions confirming that bremsstrahlung is the dominant
 radio emission mechanism in the upper shell.

\subsubsection{Origin of the Gum Nebula}
Four different models have been proposed in the literature to explain
the origin and evolution of the Gum Nebula:
\begin{enumerate}
  \item A large and moderately evolved ($\sim10^6$\,yr) $\hii$
    region, i.e., a Str{\"o}mgren sphere excited by
    $\zeta$\,Puppis and $\gamma^2$\,Velorum (\citealt{Gum1956}; 
    \citealt{Beuermann1973}).
  \item An old ($>1\,$Myr) supernova remnant that has now cooled and
    whose shell is subsequently being ionised by the early type stars
    in the interior (\citealt{Brandt1971}; \citealt{Alexander1971}).
  \item A stellar wind bubble blown by $\zeta$\,Puppis
    with help from $\gamma^2$\,Velorum and the Vela OB2-association
    (\citealt{Reynolds1976b}; \citealt{Weaver1977}).
  \item A supershell resulting from the combination of multiple
    supernova explosions and photoionising effects powered by a single
    stellar association \citep{Reynoso1997}.
\end{enumerate}
Any successful model must explain the thin {\it ionised} shell
($R/dr\sim15)$, low expansion velocity ($\sim10\,\kms$) 
and optical spectra consistent with low
excitation conditions (\citealt{Srinivasan1987};
\citealt{Sahu1993}). Classical Str{\"o}mgren sphere $\hii$ 
regions expand at approximately the observed velocity ($\sim4\,\kms$
\citealt{Lasker1966}) but do not produce a shell structure. A scaled
version of the supernova model of \citet{Chevalier1974} can produce a
bubble of the correct size, but we would then expect to see
significant radio synchrotron emission from the edge of the nebula and
this is not detected in observations to date \citep{Haslam1982}. The
old supernova remnant model also predicts that 
the cavity should be filled with $T_e\approx40,000$\,K electrons
giving rise to soft X-ray emissions. \citet{Leahy1992} detected X-ray
emitting plasma with $T_e\approx 6\times10^5$\,K towards the interior
of the Gum Nebula, but we note that this could also be explained by
the wind-blown-bubble model of \citet{Weaver1977}. The
wind-blown-bubble model also naturally explains the ionised shell
structure. 

The consensus in the literature to date favours the old SNR model of
the nebula, however, this is not universally accepted (e.g.,
\citealt{Choudhury2009}; \citealt{Urquhart2009}). 

\subsection{This work}
One way to differentiate between models of the Gum Nebula is to
examine the density profile at the edge of the shell and the effect
the nebula has on the magnetic field of the ISM. Supernovae and
wind-blown-bubbles drive strong shocks into the ISM, compressing the
gas at their leading edge. At the same time the gas inside the nebula
may be ionised by the passing shock-front (in the supernova case) or
by the central stars (in the case of wind-blown-bubbles) leading to a
corresponding increase in electron density and magnetic field strength.
Non-radiative shocks (for example in young supernovae less than
$\sim20000$ years old) expand adiabatically and we would expect to see
a density compression factor $X\lesssim4$ at the edge of the shell.
If the swept-up-shell has begun to cool radiatively (e.g.,
for snow-plough phase supernova remnants older than $\sim20000$ years)
then $X$ can be much greater - up to several hundreds. Alternatively,
if the bubble is due to a slow ionisation front moving into the
medium, we would expect little compression and would measure
$X\approx1$.

The expansion of the bubble into the ISM should also imprint a clear
signature on the Galactic magnetic field. The total field can be
visualised as a superposition of an ordered large-scale component and
a random small-scale component. The field lines are frozen into the
gas, hence compression at the bubble edge can lead to an amplification
of the field parallel the shock front. Faraday rotation is an especially
sensitive probe of the field strength along 
the line of sight and this amplification is best observed as a rotation
measure enhancement towards the limb of the shell. RMs also constitute
an excellent probe of turbulence in the ISM. Unresolved random motions
in the ionised gas can produce fluctuations in the random field that
increase the scatter between adjacent RM samples and depolarise
diffuse background polarised emission.

In this work we combine point source measurements of rotation measures
from background radio-galaxies, emission measures (EMs) from
$\halpha$ images and polarised 2.3\,GHz radio-continuum data to build a
self-consistent picture of the Gum Nebula. Using a simple geometric
model we derive the ambient electron density and magnetic field
strength. We fit for the compression factor in the shell, probe the geometry of
the ordered Galactic field and shed light on the likely origin of the nebula. 

\section{Datasets and Images}\label{sec:datasets}
We draw on data from several publicly available sky surveys. We make
use of the \citet{Taylor2009} RM catalogue, which is derived from the
1.4\,GHz NRAO VLA Sky Survey (NVSS, \citealt{Condon1996}). We estimate
emission measures using the Southern $\halpha$ Sky Survey
(SHASSA, \citealt{Gaustad2001,Finkbeiner2003}), dispersion measures
(DMs) from the Australia National Telescope Facility Pulsar
Catalogue\footnote{http://www.atnf.csiro.au/research/pulsar/psrcat,
  V1.49, August-2014}
(APC, \citealt{Manchester2005}) and we examine the polarisation
properties of the 2.3\,GHz radio-continuum maps from the S-band Parkes
All Sky Survey (S-PASS, \citealt{Carretti2011, Carretti2013a}). Below
we introduce each of the surveys and describe the processing necessary
to isolate the Gum Nebula from contaminating data.

\subsection{$\halpha$ emission}\label{sec:datasets_ha}
The $n=3-2$ Balmer series $\halpha$ recombination transition of
neutral atomic hydrogen is commonly used to derive EMs of ionised gas
in the interstellar medium. EM is directly related to the
electron density $n_e$ via ${\rm EM}=\int_{0}^{\infty}n_e^2\,dl$,
meaning that the intensity of $\halpha$ emission can be used to
estimate the line-of-sight electron density (see $\S\ref{sec:ne}$ for
a full explanation). The Southern H-Alpha Sky Survey Atlas
\citep{Gaustad2001} currently provides the highest spatial resolution
($\theta_{\rm FWHM}=6'$) coverage of the whole Gum Nebula in the
$\halpha$ emission line. We use the reprocessed SHASSA data published
by  \citet{Finkbeiner2003}, who 
subtracted point-source emission from stars, corrected for imaging
artifacts and calibrated the amplitude scale to the stable zero-point
of the Wisconsin H-Alpha Mapper (WHAM, \citealt{Haffner2003})
survey. The $\halpha$ image of the Gum Nebula is presented in
Fig.~\ref{fig:gum_summary_halpha}. The nebula describes a roughly
circular shell of emission centred on the 
Galactic Plane. Below the mid-plane the structure of the $\halpha$
data is very complicated, displaying arcs and filaments associated
with overlapping $\hii$~regions, SNR and other shells. The lower
border of the Gum Nebula appears tenuous.  Above latitudes
$b>5\degrees$ the structure of the Gum Nebula is much less confused.
The only obvious contaminating feature is the Antlia SNR
\citep{McCullough2002}, which overlaps at
$(l,b)\approx(268\degrees,\,+12\degrees)$. The northern arc of the Gum
Nebula is particularly prominent, showing a sharp edge and a
shell-like structure of width $\sim2\degrees$.

\subsubsection{Extinction correction}\label{sec:ha_extinction}
$\halpha$ emission is affected by extinction due to intervening dust
along the line of sight, characterised by the optical depth $\tau$. If
all of the dust responsible for the extinction is in the 
foreground, then the intrinsic intensity $I_{\rm H_{\alpha}}$ is reduced by 
a factor $e^{\tau}$ to give the observed intensity $I_{\rm H_{\alpha},obs}$. 
Since the location of the dust is unknown, the value $I_{\rm
  H_{\alpha}}\,=\,I_{\rm H_{\alpha},obs}\,e^{\tau}$ can be
considered a lower limit on the intensity (i.e., the maximum
correction possible) and that of $I_{\rm H_{\alpha}}\,=\,I_{\rm
H_{\alpha},obs}$ an upper limit. If the dust is uniformly mixed
with the source then the intrinsic intensity is given by $I_{\rm
  H_{\alpha}}\,=\,I_{\rm H_{\alpha},obs}\,\tau\,/\,(1-e^{-\tau})$
\citep{Reynolds1976a}.

In practice $\tau$ may be determined from the extinction observed in
the optical band, as it is related to the $E_{B-V}$ colour by
$\tau\,=2.44\times E_{B-V}$ \citep{Finkbeiner2003}. We corrected the
$\halpha$ data for extinction using the $E_{B-V}$ map created by
\citet{Schlegel1998} from the COBE (Cosmic Background
Explorer) and IRAS (Infrared Astronomical Satellite) surveys. These
$E_{B-V}$ maps provide an estimate of the total column 
of dust in the Galaxy along the line-of-sight, however, towards higher
latitudes it is reasonable to assume that most of the dust is
nearby. Dust in the plane has a scale-height of $\sim130\,$pc
\citep{Drimmel2001} and at a latitude of $b=10\degrees$, sight-lines
exit the dusty disk at a distance of $\sim800\,$pc.

Upon inspection, images of $I_{\rm H_{\alpha}}$ produced assuming all
dust is in front of the Gum Nebula appear over-corrected for prominent
dust features. For example, a filament in the $E_{B-V}$ map at
($l,b$)\,=\,($250.4\degrees,14.5\degrees$) and a circular feature at
($l,b$)\,=\,($257.0\degrees,11.9\degrees$) turn from absorption
features in $I_{\rm H_{\alpha},obs}$ to emission features in $I_{\rm
  H_{\alpha}}$. Thus, our best estimate for  $I_{\rm H_{\alpha}}$
assumes that the dust is uniformly mixed with the $\halpha$ emitting 
gas. At latitudes of $b>5\degrees$ the values of $\tau$ range over
$0.20<\tau<0.93$, corresponding to corrections of
$1.1<\tau/(1-e^{-\tau})<1.5$. \citet{Reynolds1976a} found $\tau=0.15$
towards $\zeta\,$Puppis, corresponding to $A_v=0.19$. This lower value of
optical depth is consistent if we consider that the star lies just
inside the front face of the Gum nebula. 

\subsubsection{Galactic background} 
Large-area $\halpha$ images show that emission from discrete Galactic
objects (e.g., $\hii$~regions and supernova remnants) is superimposed
on a diffuse background that rises to a peak at the Galactic
mid-plane. As seen in Fig.~\ref{fig:gum_summary_halpha}
this is a particular problem for the Gum Nebula due to its large
angular size and position straddling the plane. We have isolated the
$\halpha$ emission from the nebula by estimating and subtracting a
diffuse emission profile as a function of Galactic latitude. To
calculate the profile we identified and masked-out all
foreground objects in the image, took the minimum of the pixels in the
longitudinal direction and smoothed the resultant profile to a
resolution of $\sim1.6\degrees$. We initially created a  
background-corrected $\halpha$ map by subtracting a scaled
version of this profile from each column of pixels in the original
image. This simple scheme assumes that the background emission is
constant with longitude across the $36\degrees$ nebula;
clearly not the case since $\halpha$ emission in the right
hemisphere of the nebula is over-subtracted using this method. To
further correct the background gradient, we fit an additional
polynomial surface of order 2 to the residual large-scale
emission. After background correction, the brightness of the Gum
Nebula's shell varies between
30\,R and 170\,R away from the mid-plane
(1\,Rayleigh\,=$10^6/4\pi$\,photons~s$^{-1}$\,cm$^{-2}$\,sr$^{-1}$),
compared to a mean background level of 2\,--\,7\,R. These values are
comparable with previous estimates made using pointed spectral
observations capable of separating the Galactic and Gum Nebula
components in velocity space \citep{Reynolds1976b}.
The background-corrected $\halpha$ data are used in $\S\ref{sec:ne}$
to estimate $n_e$ along the line of sight.

\subsubsection{Uncertainties}
The formal uncertainty in the $\halpha$ intensity is a quadrature sum
of the intrinsic measurement uncertainty $\sigma(I_{\rm
  H_{\alpha}})\approx0.3\,{\rm R}$, and uncertainty due to the
extinction correction $\sigma_{\rm dust}$. As we do not explicitly
know where the dust lies along the line of sight, we assume the worst
case scenario and set the error to the likely range of correction values,
typically $\sigma_{\rm dust}\approx0.25$. For values of
$I_{H\,\alpha}$ observed towards the northern arc of the Gum Nebula
the absolute uncertainty is order 10 percent, or
$\sigma(I_{H\,\alpha})\approx12\,{\rm R}$ in the shell.

\subsection{Rotation Measures}
\begin{figure*}
  \centering
  \includegraphics[width=18.0cm, angle=0, trim=0 0 0 0]{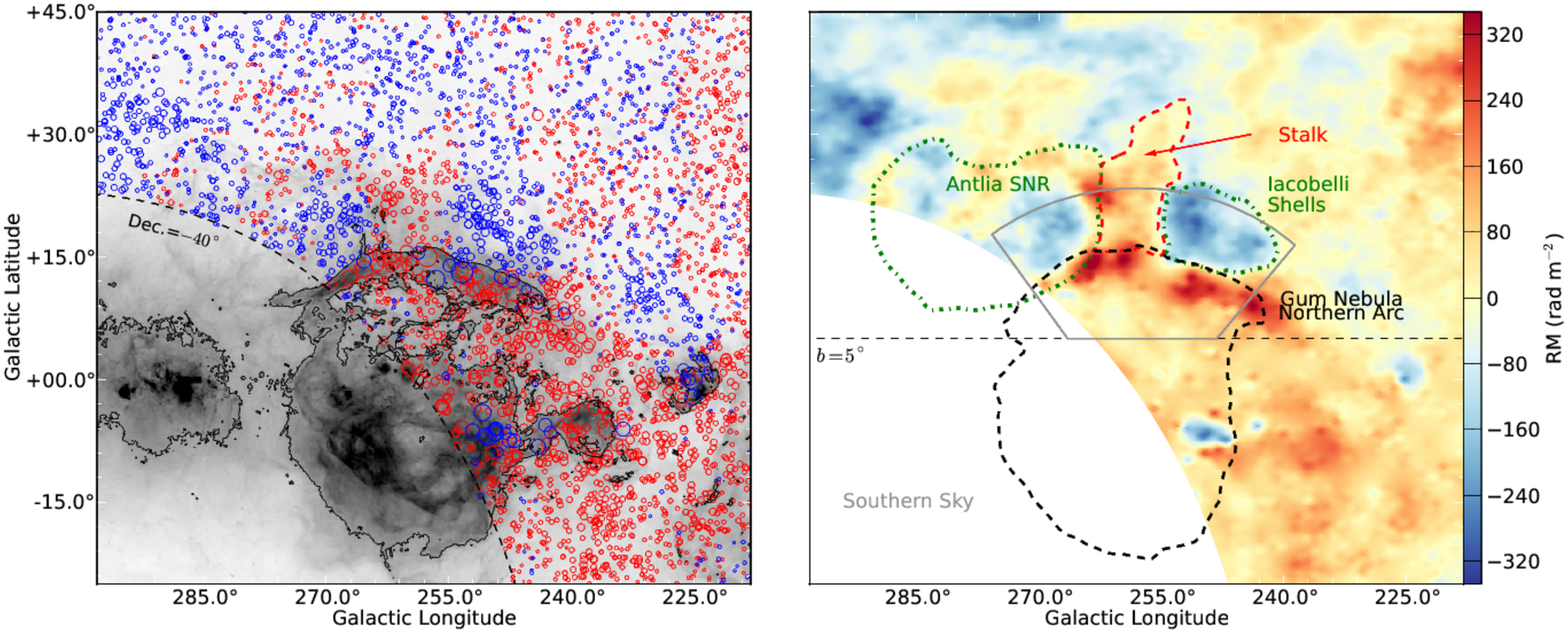}
  \caption{\small {\it Left}: The RM-catalogue of \citet{Taylor2009}
    plotted over the $\halpha$ map of the Gum Nebula
    \citep{Finkbeiner2003}. Red circles indicate positive RMs, while
    blue indicate negative and their diameter is proportional to
    $|{\rm RM}|$. The solid $\halpha$ contour at a level of 
    25\,R defines the outline of the nebula. {\it Right}: Prominent
    RM-features are easier to visualise in the map produced by
    \citet{Oppermann2012} using the \citet{Taylor2009}
    catalogue. Polygons and lines annotate significant features. We
    restrict our analysis to the upper arc region, inside the solid
    grey line.}
  \label{fig:RM_gum_overview}
\end{figure*}
\begin{figure*}
  \centering
  \includegraphics[width=18.0cm, angle=0, trim=0 0 0 0]{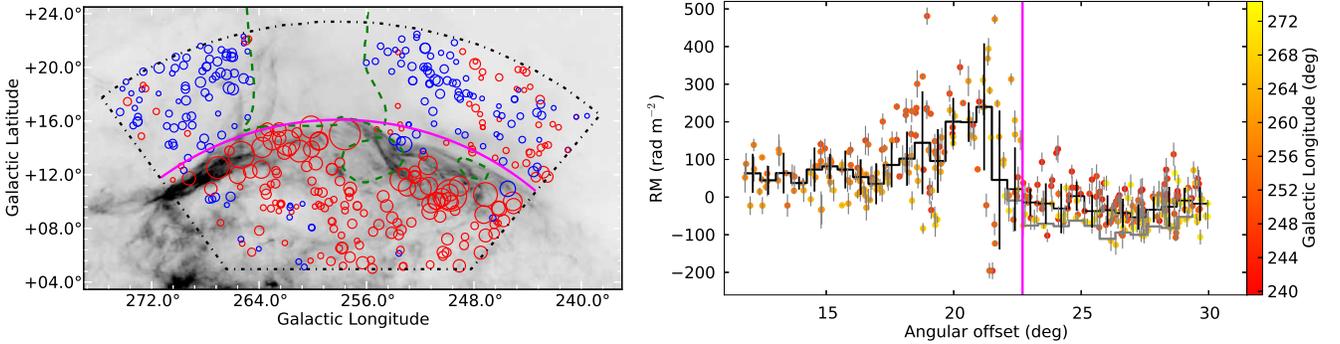}
  \caption{\small {\it Left}: Positions of polarised
    extragalactic sources selected for analysis from the
    \citet{Taylor2009} catalogue. Red and 
    blue circles indicate positive and negative RMs, respectively, and
    the radius of each circle is scaled to the absolute value of the
    RM. The green dashed lines enclose regions where sources have been
    excised from the sample because they likely probe a contaminating
    region, or are obvious outliers ($>3\sigma$). 
    {\it Right:} RMs plotted as function of angular offset from the
    centre of the nebula. The formal uncertainties in the catalogue
    are represented by vertical grey lines (scaled $\times2$ for
    clarity) and each point is colour-coded for Galactic
    Longitude. The solid black histogram represents a median-binned version
    of the plot, with bin-sizes of $0.6\degrees$. The solid grey
    histogram illustrates the level of the RMs exterior to the nebula
    before the discrete object correction was applied. Vertical black
    error-bars illustrate the $1\sigma$ scatter in each bin. The
    magenta line in both panels indicates the outer boundary of the
    nebula. In both panels the RMs are plotted assuming a flat
    background, i.e., no correction has been made for a large-scale
    RM-gradient due to the Galaxy in the background (or
    foreground). The effect of subtracting different model backgrounds
    is illustrated in Fig.~\ref{fig:RM_profile_bgs} and discussed in
    $\S\ref{sec:RMs_selected}$.}  
  \label{fig:RM_selected}
\end{figure*}
Rotation measures of polarised background radio-galaxies provide a
convenient method of measuring the line-of-sight magnetic field
$B_{||}$ in local Galactic structures, if the electron density $n_e$
and the distribution of the ionised gas are known. Each
extragalactic point source is effectively at infinity and the measured
difference between RMs of adjacent sources is dominated by local
changes in $B_{||}$ or $n_e$ along the line-of-sight path. The
effective RM contribution of a Galactic $\hii$ region, for  
example, can be found by measuring the RMs of radio-galaxies behind the
H{\scriptsize II} region and subtracting an average off-source RM,
determined from the radio-galaxies in the surrounding sky (e.g.,
\citealt{Harvey-Smith2011}).

At the present time the best sampled and most accurate large-scale
RM-grid covering a large part of the Gum Nebula is the catalogue
created by \citet{Taylor2009} from the NVSS \citep{Condon1996}. NVSS
radio-continuum observations were conducted on the Very Large Array 
(VLA) at a frequency of 1.4\,GHz and extend south to a declination of
$-40\degrees$. The original survey combined simultaneous snapshot
observations in two 42\,MHz-wide bands (1364.9\,MHz and 1435.1\,MHz)
into a multi-frequency synthesis image of the sky in Stokes~{\it I},
{\it Q}, {\it U} and~{\it V}. \citet{Taylor2009} reprocessed the NVSS
visibility data into individual images of the bands and calculated
two-channel rotation measures for 37,543 polarised sources.

\subsubsection{RMs through the Gum Nebula}\label{sec:RMs_selected}
Fig.~\ref{fig:RM_gum_overview}\,-\,{\it left} presents RMs from the
\citet{Taylor2009} catalogue over-plotted on the  $\halpha$ image of
the Gum Nebula. Although the lower left of the nebula is not
covered by the NVSS, there are RM measurements towards the
upper right with source densities varying between 1.2 and
6.6\,deg$^{-2}$ \citep{Stil2007}. It is easier to visualise
RM-features in the smoothed map created by \citet{Oppermann2012} mainly from
the \citet{Taylor2009} catalogue  and plotted in
Fig.~\ref{fig:RM_gum_overview}\,-\,{\it right}.  Using the $\halpha$
and radio continuum maps as a guide, we identify structures in the RM
map likely associated with Galactic objects. The RM-signature of the
Gum Nebula is clearly different from the background and matches the 
morphology of the excited hydrogen gas well. The distinctive upper
arc of the nebula displays consistently positive RMs, which decrease in
magnitude towards the geometric centre, reminiscent of a limb-brightened
shell. The net positive RM signal in the upper Gum Nebula implies a
coherent magnetic field on scales of $\sim260$\,pc, the projected
diameter of the nebula at the adopted distance of $d=450$\,pc. The
region along
the Galactic mid-plane ($|b|\lesssim5\degrees$) is highly confused,
containing several $\hii$~regions and supernova  remnants (see
Fig.~\ref{fig:gum_summary_halpha}), while the lower part of the nebula
also contains the IRAS Vela Shell, which may be a separate foreground
object. In contrast, the upper part of the Gum Nebula appears relatively free of
obscuring objects and we focus on this region for the remainder of the
paper. The grey wedge-shaped box in Fig.~\ref{fig:RM_gum_overview}
outlines the data selected for analysis. The selected region includes
the upper arc of the nebula, the  interior above $b=5\degrees$ and a
section of off-source RMs outside of the nebula's
border.

\subsubsection{Isolating the RM-signature of the Gum Nebula}\label{sec:RM_isolate}
Any analysis of the RM data relies on isolating the
RM-signature of the Gum Nebula from discrete regions of magneto-ionic
material along the line of sight (e.g., overlapping supernova remnants
or $\hii$ regions) and from the bulk of the Galaxy in the
background. We have identified three other sets of discrete Galactic objects
towards the upper Gum Nebula that have Faraday rotation signatures in
the \citet{Taylor2009} RM-catalogue. The Antlia SNR
(\citealt{McCullough2002}, annotated in
Fig.~\ref{fig:RM_gum_overview}) is adjacent to the nebula on the upper
left. The edge of the SNR is traced by sources with RM values
$10-15\,\rmtwo$ more positive than their surroundings. This excess RM 
signal is comparable to the formal error in the catalogue
($\sim12\,\rmtwo$) and is negligible compared to RMs through the rim of
the Gum Nebula ($\sim300\,\rmtwo$). RMs through the interior of the
Antlia SNR are consistent with the background, except for a patch
directly bordering the Gum Nebula, which is more negative than the
large scale background by approximately $-30.0\,\rmtwo$. This patch
lies inside our selection box so we subtract this offset from RMs
inside the patch to correct the catalogue. The 
second obvious feature in the data is a pair of shells discovered by
\citet{Iacobelli2014} in 2.3\,GHz radio continuum data, seen in
polarisation and lying to the upper-right of the Gum Nebula. The
border of the shells seen in 
the radio data corresponds exactly to the morphology of a negative patch
of RMs in Fig.~\ref{fig:RM_gum_overview}. We again correct the
catalogue by subtracting the median background offset
($-77.2\,\rmtwo$) from RMs inside the shell boundaries. The final feature 
of note is a `stalk' of positive RMs extending from the centre of
the upper arc to higher Galactic latitudes. The `stalk' has a
counterpart in $\hi$ emission identified by \citet{Reynoso1997}, lies
above a hole in the $\halpha$ image and is hypothesised by the authors to
be a `blowout' in the shell wall leading to ionised gas streaming into the
Galactic halo. Modelling and subtracting the signature of this feature
is beyond the scope of this work, so we simply mask off the RM data
within its boundary. Fig.~\ref{fig:RM_selected}\,-\,{\it left} presents
the corrected RM catalogue within the selection box, plotted over the
$\halpha$ image. The azimuthally-averaged RM-profile is shown in
Fig.~\ref{fig:RM_selected}\,-\,{\it right}). The black histogram shows a
version of the profile binned in $0.6\degree$ increments. Outside of
the nebula border (offsets $\gtrsim22\degrees$) the RMs are relatively
constant but rise rapidly to a peak just inside the border. At smaller
offsets the RM values fall slowly, approaching an interior level that
is higher than the background. The grey histogram, shows the same binned
profile prior to correcting for discrete contaminating sources. Note
that the RM data shown here have not yet been corrected for
large-scale gradients due to diffuse thermal electrons distributed
throughout the bulk of the Galaxy in the background.

\begin{figure}
  \centering
  \includegraphics[width=7.5cm, angle=0, trim=0 0 0 0]{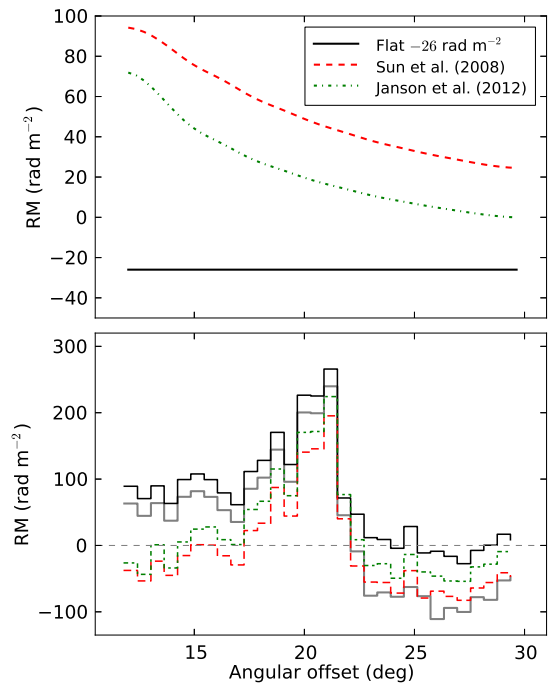}
  \caption{\small {\it Top}: Sample profiles at $l=259\degrees$ from
    the models of \citet{Sun2008} and \citet{Jansson2012} illustrating
    the predicted RM-gradient with Galactic latitude. {\it Bottom}:
    The light-grey histogram is the azimuthally averaged RM-profile of the
    Gum Nebula uncorrected for discrete contaminating sources or
    large-scale RM background (also plotted in
    Fig.~\ref{fig:RM_selected}). The black histogram is 
    the equivalent with discrete corrections applied and a flat
    background of $-26.4\,\rmtwo$ subtracted. The red (dashed) and green
    (dot-dashed) histograms illustrate the effect of subtracting the 2-D
    \citet{Sun2008} and \citet{Jansson2012} RM-models,
    respectively, from each point source select from the
    \citet{Taylor2009} catalogue. Note that the discrete corrections
    have previously been applied. In both cases the large-scale gradient
    correction acts to decrease the RMs in the interior of the 
    nebula relative to the exterior.}
  \label{fig:RM_profile_bgs}
\end{figure}

The distribution of electrons within the Galaxy and the strength, and
geometry, of the ordered Galactic magnetic field result in a unique
pattern of RMs over the whole sky. In the all-sky RM map compiled by
\citet{Oppermann2012} the dominant signal is quadrupolar in shape,
with negative RMs above and positive below the Galactic mid-plane in
the vicinity of the Gum Nebula. In recent years several authors have 
modelled the ordered Galactic field by combining data from
extra-galactic RMs and radio-synchrotron emission
\citep{Sun2008,Jansson2009,Sun2010,Mao2010,Jaffe2010,vanEck2011,Jansson2012},
and explain the pattern as being due to the toroidal field in
the halo. Within the Galactic disk the magnetic field and
thermal electron density follow the spiral arms, increasing towards
the mid-plane, leading to steep gradients in RM at low Galactic
latitudes \citep{Simard1980,Cordes2002,Gaensler2008}. The sparse sampling of the 
\citet{Taylor2009} RM-catalogue and confusion towards the mid-plane
mean that accurately removing the large-scale RM-signal due to the
Galaxy is challenging. We initially attempted to fit a 2D polynomial
surface to the off-source RMs, but this proved to be highly
unreliable in practice. Instead we consider two classes of potential
RM backgrounds. In the first case we assume a simple flat background at
the median level of the selected RMs outside the boundary of the Gum
Nebula: $-26.4\,\rmtwo$. This assumption is the simplest correction
possible and consistent with the high-latitude RM-data. However, we
know that the volume-averaged electron density decays exponentially
with height above the mid-plane \citep{Cordes2002,Gaensler2008}, thus
the RMs must decrease correspondingly. 
\citet{Sun2008} and \citet{Jansson2012} have modelled the large scale
RMs distribution of the Galaxy starting from the NE2001 electron
density distribution and applying the scale height corrections of
\citet{Gaensler2008}. Both models have similar latitude profiles,
illustrated for $l=258\degrees$ in Fig.~\ref{fig:RM_profile_bgs}\,-\,{\it
  top}. The bottom panel of Fig.~\ref{fig:RM_profile_bgs} shows the
effect of subtracting each 2-D model from the selected RM
data-points. The resulting azimuthal profile is essentially the same
for both models, but offset in RM as the models differ in their
absolute calibration. Both models act to decrease the value of RMs
towards the interior of the nebula. Neither model correctly
predicts the absolute zero-level exterior to the Gum Nebula, likely
because the best-fitting models are constrained over the whole sky and
by other, sometimes contradictory, datasets. The authors also had
limited knowledge of local contaminating objects. We adopt the
RM-models of \citet{Sun2008} and \citet{Jansson2012} as the best
available estimates of the large-scale background variation in RM and
apply offsets of $-64.0\,\rmtwo$ and~$-36.6\,\rmtwo$, respectively, so
as to correct their calibration to the zero-point exterior to the Gum
Nebula. In our analysis of the RMs through the Gum Nebula we compare
the results derived with each of these backgrounds separately.

\subsection{2.3\,GHz radio continuum}\label{sec:radio_data}
\begin{figure*}
  \centering
  \includegraphics[width=18.0cm, angle=0, trim=0 0 0 0]{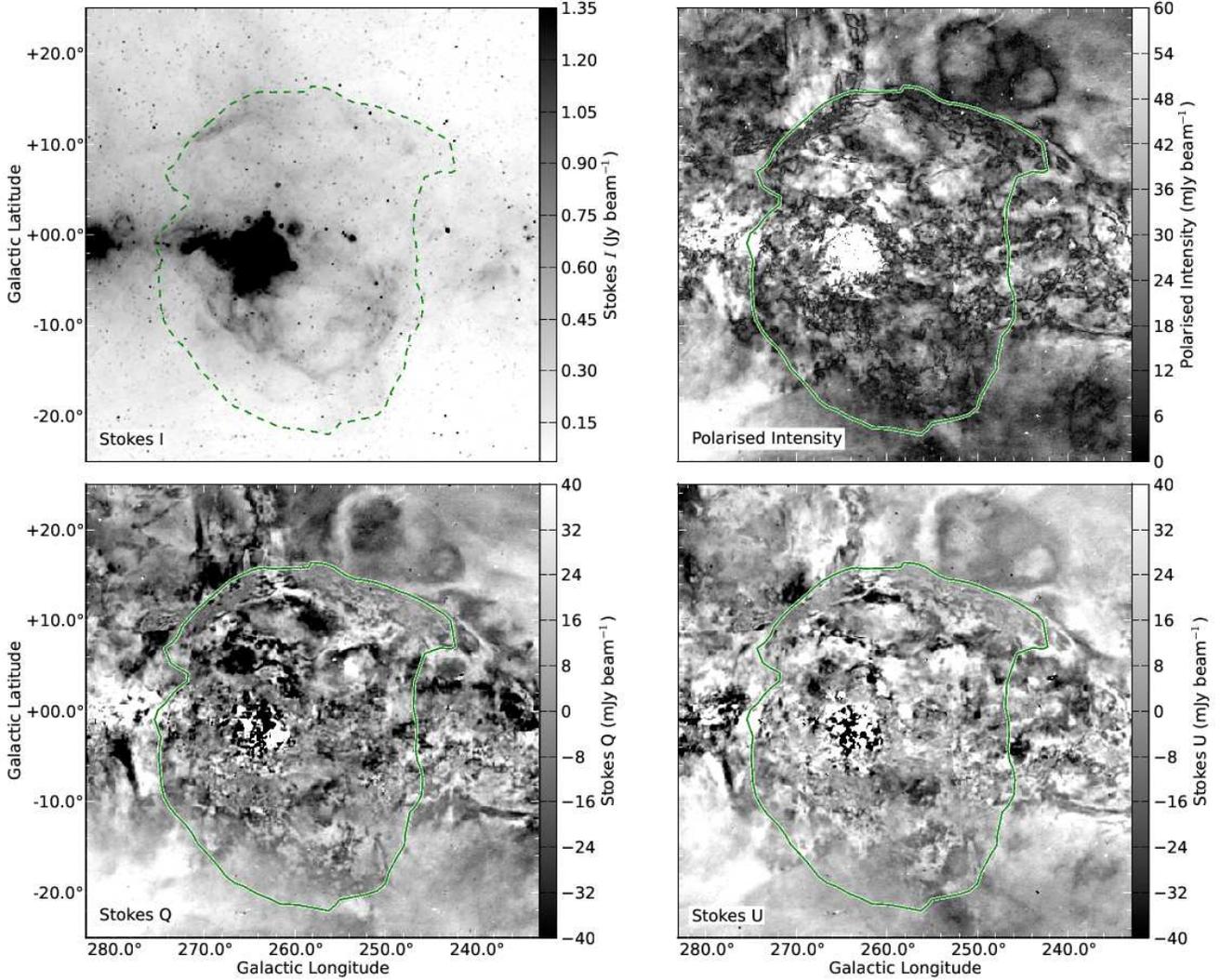}
  \caption{\small 2.3\,GHz radio-continuum maps of the Gum Nebula
    from the S-PASS project. The data are calibrated in Janskys and
    may be converted to a main-beam brightness temperature scale in
    Kelvin by multiplying by 0.55. The border of the nebula traced by
    the green line is the same as in Fig.~\ref{fig:RM_gum_overview}. }
  \label{fig:radio_continuum}
\end{figure*}
Radio continuum at centimetre wavelengths traces gas emitting via
both synchrotron and thermal processes. If information on the
polarisation state of the radiation is available, analysis of the
Stokes~{\it Q} and~{\it U} parameters can constrain conditions in the
gas along the line-of-sight, e.g., depolarisation due to a fluctuating
component of the magnetic field.

The S-band Polarisation All Sky Survey (S-PASS) has imaged the entire
southern sky (${\rm Dec.}<-1\degrees$) in polarisation at a frequency
of 2.3\,GHz.  The observations have been conducted with the Parkes
Radio Telescope, NSW Australia, a 64-m telescope operated by CSIRO
Astronomy and Space Science. A description of S-PASS observations and
analysis is given in \citet{Carretti2010} and \citet{Carretti2013a}. Here
we report a summary of the main details. The standard S-band receiver
of the observatory ({\it Galileo}) was used with a system temperature
$T_{\rm sys}=20$\,K, beam width FHWM$=8.9'$ at 2300\,MHz and a
circular polarisation front--end ideal for linear polarisation
measurements with a single-dish telescope. Data have been detected
with the Digital Filter Banks mark 3 (DFB3) with full Stokes
capabilities recording the two autocorrelation (RR and LL) and the
complex cross-correlation products of the two circular polarisations
(RR, LL, LR, RL$^*$). Flux calibration was done with PKS\,B1934$-$638,
secondary calibration with PKS\,B0407$-$658 and polarisation
calibration with PKS\,B0043-424. Data were binned in 8\,MHz channels
and, after RFI flagging, 23~sub-bands were used, covering the ranges
2176-2216 and 2256-2400\,MHz, for an effective central frequency of
2307\,MHz and bandwidth of 184\,MHz.

The observing strategy is based on long azimuth scans taken towards
the East and the West at the elevation of the south celestial pole at
Parkes (EL$=33\degrees$) to realise absolute polarisation calibration
of the data. Final maps are convolved to a beam of
FWHM$=10.75'$. Stokes~{\it I}, {\it Q}, and~{\it U} sensitivity is
better than $1.0\,\mjbm$ per beam-sized pixel everywhere in the
covered area. Details of scanning strategy, map-making, and final maps
obtained by binning all frequency channels are presented in
\citet{Carretti2010} and Carretti et al. (2015, in preparation). The
confusion limit is 6\,mJy in Stokes~{\it I} \citep{Carretti2013b} and
much lower in polarisation (average polarisation fraction in compact
sources is lower than 2\,percent, \citealt{Tucci2004}). The
instrumental polarisation leakage is 0.4\,percent on-axis
\citep{Carretti2010} and less than 1.5\,percent off-axis. For diffuse
emission, the latter is generally not important because of
cancellation effects at scales larger than the beam (e.g.,
\citealt{Carretti2004, ODea2007}).

\subsubsection{The Polarised Signature of the Gum Nebula}
Fig.~\ref{fig:radio_continuum} presents the 2.3\,GHz radio-continuum
image of the Gum Nebula in Stokes~{\it I}, {\it Q}, {\it U} and
polarised intensity $P$. The morphology of the nebula in total
intensity is broadly similar to the $\halpha$ map presented in
Fig.~\ref{fig:gum_summary_halpha}, implying that the radio- and
optical-emission are coming from the same gas. When viewed
in {\it P}, {\it Q} and {\it U}, the upper shell of the nebula is seen
to depolarise background emission in a $\sim2\degree$ wide arc. This
band of depolarisation is set against the smooth Galactic background,
visible in the upper-right quadrant of the image, above
$b=12\degrees$. At high latitudes the background is also depolarised
by two thin shells (\citealt{Iacobelli2014}, see
$\S\ref{sec:RMs_selected}$ and Fig.\,\ref{fig:RM_gum_overview})
and by the Antlia supernova remnant in the upper-left quadrant. The
shell of the Antlia SNR is similarly 
characterised by a band of depolarisation that overlaps the Gum Nebula
at $(l,b)\approx(268\degrees,+13\degrees)$. The interiors of the Gum Nebula
and Antlia SNR appear fractured, exhibiting patches of homogeneous
polarised intensity interspersed with depolarised `canals'. The Vela
supernova remnant at $(l,b)\,=\,(267\degrees,-3\degrees)$ is the
brightest object in the field, ($I$ \& $P$), while the rest of the
Galactic plane is seen as a mix of polarised foreground and
depolarised background emission.

\subsection{Pulsars}
The ATNF Pulsar Catalogue (APC, \citealt{Manchester2005}) collates the
properties of more than 2300 rotation-powered pulsars and is
continually revised as new discoveries are made. Of primary interest
to us are the dispersion measures (DMs), RMs and distances to the
pulsars. These parameters can be combined with EMs and a geometric
model to derive the average electron density, filling factor and
magnetic field strength along the line-of-sight.

\begin{table*}
  \caption{Properties of selected pulsars towards the Gum Nebula.}
\begin{scriptsize}\label{tab:pulsars}
  \begin{tabular}{lcc r@{\,$\pm$\,}l r@{\,$\pm$\,}l cl}
    \tableline
    \tableline
    \rule[-0.3em]{0pt}{1.35em}(1)       & (2)    & (3)  & \multicolumn{2}{c}{(4)} & \multicolumn{2}{c}{(5)} & (6) & (7)\\
    \rule[-0.4em]{0pt}{1.45em} Name & $l$ & $b$ & \multicolumn{2}{c}{DM} & \multicolumn{2}{c}{RM} & $D$ & Notes \\
    \rule[-0.4em]{0pt}{1.45em}      & (deg.) & (deg.) & \multicolumn{2}{c}{(${\rm cm^{-3}\,pc}$)} & \multicolumn{2}{c}{(${\rm rad\,m^{-2}}$)} & (kpc) & \\
    \tableline
\multicolumn{5}{l}{\rule[-0.5em]{0pt}{1.55em}{\bf Pulsars towards the upper Gum Nebula:}}\\
\cline{1-5}
\rule[0mm]{0pt}{1.1em}J0758-1528 & 234.464 & 07.224 & 63.327 & 0.003 & 55 & 7 & 3.72 & Out \\
J0818-3049 & 249.983 & 02.908 & 133.7 & 0.2 & \multicolumn{2}{c}{~~--} & 4.17 & Gum \\
J0820-1350 & 235.890 & 12.595 & 40.938 & 0.003 & $-$1.2 & 0.4 & 1.9$^{\dagger}$ & Out \\
J0828-3417 & 253.965 & 02.561 & 52.2 & 0.6 & 59 & 3 & 0.5 & Gum \\
J0838-2621 & 248.807 & 08.981 & 116.9 & 0.1 & 86 & 13 & 4.6 & Gum \\
J0846-3533 & 257.190 & 04.710 & 94.16 & 0.11 & 144 & 8 & 1.4 & Gum \\
J0855-3331 & 256.847 & 07.517 & 86.635 & 0.016 & 165 & 10 & 1.2 & Gum \\
J0900-3144 & 256.162 & 09.486 & 75.702 & 0.010 & \multicolumn{2}{c}{~~--} & 0.8 & Gum \\
J0904-4246 & 265.075 & 02.859 & 145.8 & 0.5 & 284 & 15 & 4.4 & Gum \\
J0908-1739 & 246.119 & 19.850 & 15.888 & 0.003 & $-$31 & 4 & 0.6 & Out \\
J0912-3851 & 263.165 & 06.584 & 70 & 1 & \multicolumn{2}{c}{~~--} & 0.6 & Gum \\
J0923-31 & 259.697 & 13.003 & 72 & 20 & \multicolumn{2}{c}{~~--} & 1.0 & Gum \\
J0932-3217 & 261.277 & 14.069 & 102.1 & 0.8 & \multicolumn{2}{c}{~~--} & 3.8 & Gum \\
J0934-4154 & 268.361 & 07.411 & 113.79 & 0.16 & \multicolumn{2}{c}{~~--} & 3.2 & Gum \\
J0941-39 & 267.795 & 09.904 & 78.2 & 2.7 & \multicolumn{2}{c}{~~--} & 1.3 & Gum \\
J0945-4833 & 274.199 & 03.674 & 98.1 & 0.3 & \multicolumn{2}{c}{~~--} & 2.7 & Out \\
J0952-3839 & 268.702 & 12.033 & 167 & 3 & \multicolumn{2}{c}{~~--} & 8.4 & Gum,\,Ant \\
J0959-4809 & 275.742 & 05.418 & 92.7 & 1.2 & 50 & 6 & 3.0 & Out \\
J1000-5149 & 278.107 & 02.603 & 72.8 & 0.3 & 46 & 9 & 2.3 & Out \\
J1003-4747 & 276.037 & 06.117 & 98.1 & 1.2 & 18 & 4 & 3.4 & Out \\
J1012-2337 & 262.131 & 26.377 & 22.51 & 0.09 & 52 & 9 & 1.3 & Out \\
J1032-5206 & 282.354 & 05.128 & 139 & 4 & \multicolumn{2}{c}{~~--} & 4.3 & Out \\
J1034-3224 & 272.050 & 22.117 & 50.75 & 0.08 & $-$8 & 1 & 4.7 & Ant \\
J1036-4926 & 281.518 & 07.727 & 136.529 & 0.010 & $-$11 & 6 & 8.7 & Out \\
J1045-4509 & 280.851 & 12.254 & 58.166 & 0.001 & 92 & 1 & 0.23$^{\dagger}$ & Ant \\
J1057-4754 & 284.007 & 10.739 & 60 & 8 & \multicolumn{2}{c}{~~--} & 3.0 & Ant \\
J1105-43 & 283.511 & 14.886 & 38.000 & 0.001 & \multicolumn{2}{c}{~~--} & 2.2 & Ant \\
\multicolumn{5}{l}{\rule[-0.5em]{0pt}{1.55em}{\bf Pulsars below $b=2\degrees$ with accurate distances:}}\\
\cline{1-5}
\rule[0mm]{0pt}{1.1em}J1001-5507 & 280.226 & 00.085 & 130.32 & 0.17 & 297 & 18 & 0.3$^{\dagger}$ & Out \\
J0737-3039B & 245.236 & $-$4.505 & 48.920 & 0.005 & 112.3 & 1.5 & 1.1$^{\dagger}$ & Out \\
J0738-4042 & 254.194 & $-$9.192 & 160.8 & 0.7 & 12.1 & 0.6 & 1.6$^{\dagger}$ & Gum \\
J0742-2822 & 243.773 & $-$2.444 & 73.782 & 0.002 & 149.9 & 0.1 & 2.0$^{\dagger}$ & Out \\
J0835-4510 & 263.552 & $-$2.787 & 67.99 & 0.01 & 31.4 & 0.1 & 0.28$^{\dagger}$ & Gum \\
J0837-4135 & 260.904 & $-$0.336 & 147.29 & 0.07 & 135.8 & 0.3 & 1.5$^{\dagger}$ & Gum \\
J0908-4913 & 270.266 & $-$1.019 & 180.37 & 0.04 & 10.0 & 1.6 & 1.0$^{\dagger}$ & Gum \\
J0942-5552 & 278.571 & $-$2.230 & 180.2 & 0.5 & -61.9 & 0.2 & 0.3$^{\dagger}$ & Out \\
    \tableline
  \end{tabular}
\end{scriptsize}
\tablecomments{This table presents the properties of the pulsars
  displayed in Figure~\ref{fig:pulsar_map}. Pulsars marked with a
  $\dagger$ in column (6) have independent distance measurements;
  other entries default to the DM-derived distance. The code in column (7)
  notes whether the pulsar falls on a sightline towards the Gum Nebula
  (Gum), towards the Antlia SNR (Ant) or outside of the border of
  either object (Out).} 
\end{table*}

\begin{figure*}
  \centering
  \includegraphics[width=18.0cm, angle=0, trim=0 0 0 0]{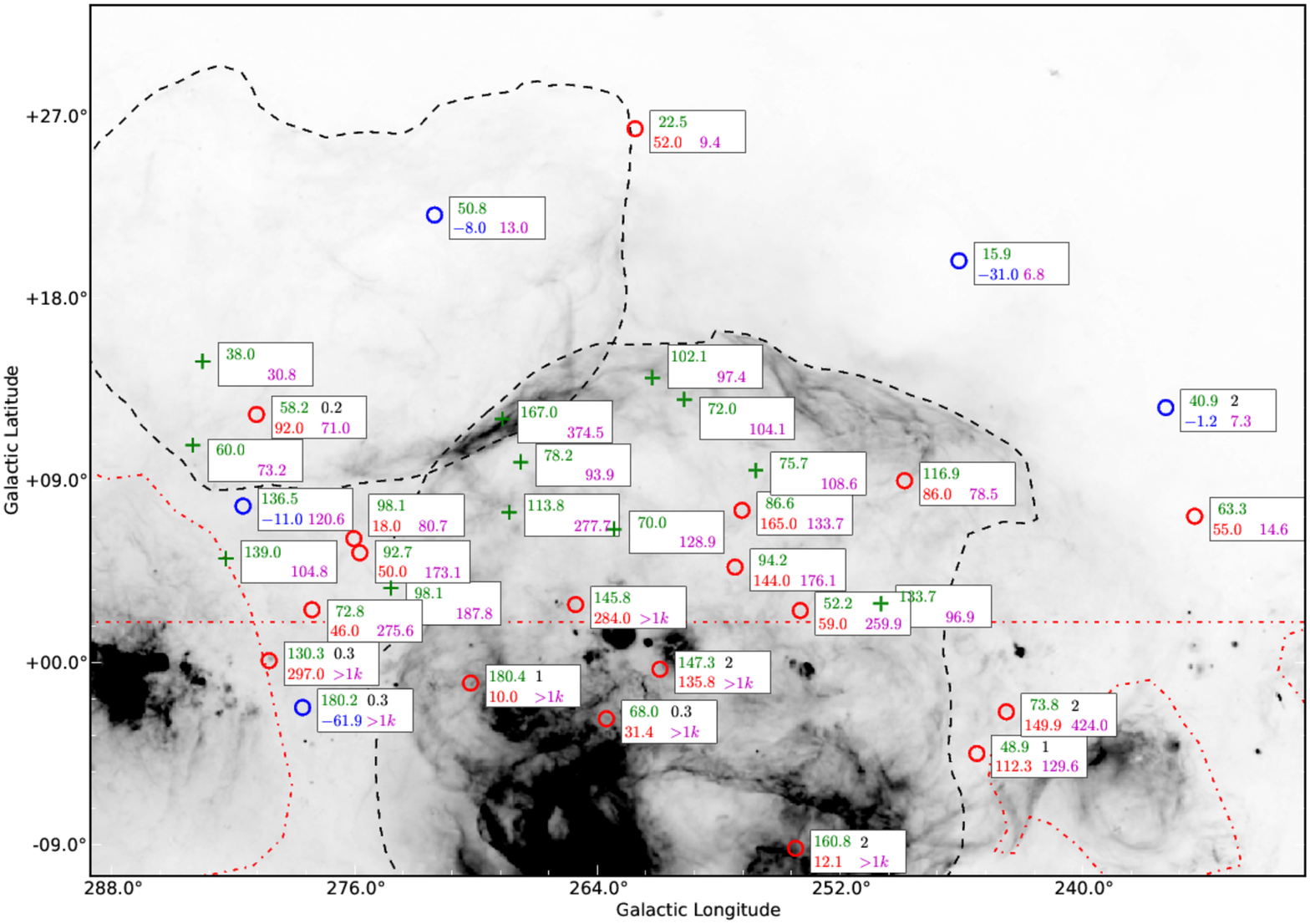}
  \caption{\small Selection of pulsars with known DM values towards
    the Gum Nebula. All pulsars with $b>2\degrees$ are shown
    alongside a selection of pulsars with accurate distances below
    $b=2\degrees$. Red dot-dashed lines outline confusing parts of the
    Galactic plane and pulsars in these regions have been
    omitted. `$\circ$' symbols represent pulsars with both DM and RM
    measurements, and `$+$' symbols pulsars with only DM
    measurements. Each pulsar is annotated with its DM in
    $\pccmmthree$ (top-left number, coloured green), RM in $\rmtwo$
    (bottom-left number, coloured red or blue signifying positive or
    negative RM, respectively), independently determined distance in
    kpc, where known (top-right number, coloured black) and EM in
    $\pccmmsix$ (bottom-right number, coloured magenta).}
  \label{fig:pulsar_map}
\end{figure*}
We utilise version 1.49 of the APS, which lists 158 pulsars within a
$30\,\degree$ radius of the kinematic centre of the Gum Nebula
($l=261.0\degrees,~~~b=-2.5\degrees$; \citealt{Woermann2001}). Of
these we have chosen 35 for analysis, most of which lie within the
upper Gum region. Fig.~\ref{fig:pulsar_map} and
Table~\ref{tab:pulsars} present this sample, which contains all
pulsars above $b=2\degrees$ and a handful below, chosen because they
have accurately determined distances or lie on unconfused sight-lines
adjacent to the Gum Nebula. 

Accurate distances to pulsars are difficult to obtain:
a handful of precise values have been calculated via annual
parallaxes and these are limited to relatively nearby pulsars
($<3$\,kpc). Kinematic distances accurate to $\sim1$\,kpc can be
derived for some pulsars associated with H\,{\scriptsize I} absorption,
while the distance to pulsars located in globular clusters can be
estimated to $\sim15\%$ reliability via analysis of colour-magnitude
diagrams. Most of the pulsars detected towards the Gum Nebula default to a
distance derived from the dispersion measure. Such DM-distances 
are often highly inaccurate because they rely on a model of the Galactic
free-electron distribution \citep{Cordes2002,Taylor1993a}, which was
itself created in part using the \citet{Taylor1993b} pulsar
catalogue. Distances are particularly ill-determined towards the Gum
Nebula, which was included in the \citet{Cordes2002} model as a pair
of overlapping spheres of diameter 50\,pc. It is not clear that this
is an improvement on the older \citet{Taylor1993a} model which treated
the nebula a simple Gaussian of full width half maximum (FWHM) 50\,pc
truncated at $r=130$\,pc. The \citet{Cordes2002} model does, however,
account for the scatter broadening $\tau_{\rm sc}$ which was measured by
\citet{Mitra2001} for 40 pulsars between $250\degrees<l<290\degrees$. They
found that $\tau_{\rm sc}$ was greater than expected for a smooth
Gaussian, implying a more inhomogeneous distribution of $n_e$. Based
on the observed scattering they concluded that pulsars in the vicinity
of the Gum Nebula should be 2\,-\,3 times closer than predicted by
\citet{Taylor1993a}. 
Within the area of the Gum Nebula only four pulsars have both DM
measurements and accurate distances. The Vela pulsar (J0835$-$4510) is
known to be at a distance of $287^{+19}_{-17}$\,pc \citep{Dodson2003},
placing it just inside the front wall of the nebula. The remaining
three (J0738$-$4042, J0837$-$4135 and J0908$-$4913) lie at
distances greater than 1\,kpc (see the bottom of
Table~\ref{tab:pulsars}), behind the Gum Nebula.

\section{Analysis}\label{sec:analysis}
Our analysis aims to answer the questions: {\it What is the likely
  origin of the Gum Nebula?} and {\it What are the magnetic properties
of the nebula and how do they affect ambient conditions in this part
of the Galaxy?} To address these questions we construct a simple model
of the nebula as an ionised shell situated in the
near-field. We present the model below and explain the
maximum-likelihood method used to fit the model to RMs
on the sky. The model assumes a uniform density distribution plus a
jump in ionisation fraction from 0 to 100 percent within the shell of
the Gum Nebula, which we derive from the $\halpha$ data and include as
a prior in our fitting procedure. The  resulting fits will be
presented in $\S\ref{sec:results}$.

\subsection{Rotation Measures as Magnetic Probes}\label{sec:RM_as_probe}
Faraday rotation causes the polarisation angle of a
linearly polarised wave traversing a magnetised ionised medium to
rotate by an angle $\Delta\psi$. The change in polarisation angle is
given by
\begin{equation}
  \Delta\psi = {\rm RM}\,\lambda^2~~~{\rm rad},
\end{equation}
where RM is the rotation measure in radians\,m$^{-2}$. The observed RM
depends on the line-of-sight component of the magnetic field $B_{||}$
(in $\mu$G), the thermal electron density $n_e$ (in ${\rm cm^{-3}}$)
and the path length $dl$ (in pc) according to 
\begin{equation}\label{eqn:RM}
  {\rm RM} = 0.81~\int_{src}^{obs}n_e\,B_{||}\,dl~~~{\rm rad\,m^{-2}}.
\end{equation}
Note that the integral in Equation~\ref{eqn:RM} is taken from the
source of the polarised emission to the observer, so that a positive
RM indicates an average magnetic field pointing towards the observer.
If the ionised material along the line-of-sight contains
clumps of uniform $n_e$ threaded by the same $B_{||}$, then 
the medium is characterised by a volume filling factor
$f$. Equation~\ref{eqn:RM} becomes
\begin{equation}\label{eqn:RM_1}
  {\rm RM} = 0.81\,n_e\,B_{||}\,f\,L~~~{\rm rad\,m^{-2}},
\end{equation}
where $L$ is the total path length through the ionised medium and
$f\,L$ is known as the occupation length. $L$ can
generally be estimated from the geometry of the object under
consideration (e.g., a slab, sphere or shell).

\subsection{A Near-Field Magnetic Bubble Model}\label{sec:mag_bubble}
\begin{figure*}
  \centering
  \includegraphics[angle=90, width=18.0cm, trim=0 0 0 0]{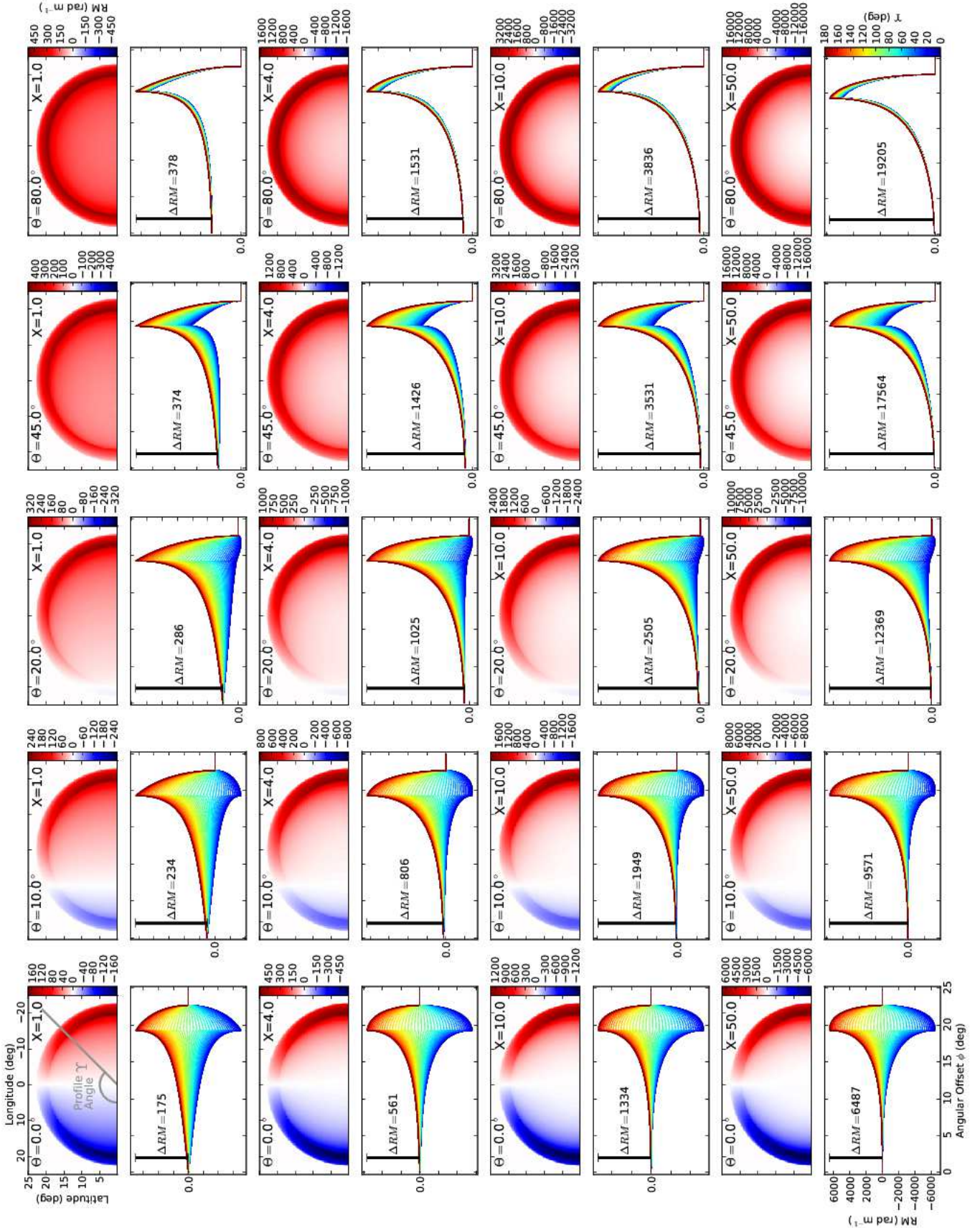}
  \caption{\small Grid of models showing how changes in the angle of
    the magnetic field $\Theta$ and the compression factor $X$ affect
    the distribution of RMs across a simple ionised shell. The
    geometry of the model is described in the Appendix (see Fig.~A\,1.) 
    and the parameters are set to:
    $D=450$\,pc, $\phi_{outer}=22.7\degrees$, $dr=25.0$\,pc,
    $n_e=1.7\,\cmmthree$, $B_0=8.6\,\mu$G and $f=0.5$. Each model is
    presented in two panels: the upper panel presents a map of RM in
    offset Galactic coordinates and the lower panel displays radial RM
    profiles extracted over a range of angles $\Upsilon$ to the
    Galactic plane ($0\degrees\leq\Upsilon\leq180\degrees$).}
  \label{fig:bubble_grid_lrg}
\end{figure*}

Models of RMs through spherical ionised shells have recently been used
to  derive magnetic properties of Galactic SNRs and $\hii$~regions, and
to probe the 
magnitude and orientation of the ordered Galactic magnetic field,
e.g. \citet{Savage2013}, \citet{Harvey-Smith2011}, \citet{Whiting2009} and
\citet{Kothes2009}. These phenomena ionise their surroundings and
illuminate the ambient magnetic field via Faraday rotation. As they
expand into the ISM they may also compress the field, imprinting a
specific signature on the rotation measures. Previous investigations
have focused on distant objects ($>1$\,kpc) whose small angular diameters 
($<5\degrees$) mean that they intercept fewer RM sight-lines compared to
the nearby Gum Nebula. Because these bubbles lie in the
far-field, their RM profiles may be integrated in azimuth under the
assumption of spherical symmetry. However, for a near-field bubble
like the Gum Nebula, the sign and shape of the RM-profile can vary
with azimuth, depending on the orientation of the ordered magnetic
field.

In a similar way to \citet{Whiting2009} and \citet{Savage2013}, we
model the Gum Nebula as a spherical ionised shell of radius $R$ and
thickness $dr$, threaded by a uniform, parallel magnetic field
$\overrightarrow{B_0}$. We assume that the electron density $n_e$ is
constant within the shell and zero elsewhere (i.e., the background has
been removed as in $\S\ref{sec:datasets_ha}$ and the electron density
in the interior of the shell is negligible). The observer is located
in the near-field and the magnetic field lines make an angle $\Theta$
to the plane of  the sky in the direction of the bubble centre. The tilt
angle\footnote{The tilt angle is the angle the magnetic field makes to
  the Galactic disk at the position of the nebula} of the magnetic
field is assumed fixed 
along the y-axis, representing the Galactic plane, and the angle
$\zeta$ describes the orientation of the sight-line to the
yz-plane. The edge of the shell subtends an angular radius of
$\phi_{\rm outer}={\rm sin}^{-1}(R/D)$, where $D$ is the distance from
the observer to the geometric  centre of the shell. A full description
of the adopted geometry can be found in Appendix~\ref{app:shell_geom},
including a detailed schematic.

If the shell is expanding supersonically into the ISM then the gas,
and hence the magnetic field, will be compressed at the external
boundary. If the expansion has slowed to sub-sonic speeds then the
gas will simply move out of the way. To model the compression (or lack
of) we assume the electron density behind the expansion front 
is given by $n_e=X\,n_0$, where $X$ is the compression factor and
$n_0$ is the electron density in the ambient medium. At each
point on the sphere the component of the magnetic field tangent to
the shell ($B_{\perp}$) is amplified by $X$ while the normal
component ($B_n$) is unaffected. The contribution to the observed
magnetic field by one hemisphere is simply the vector sum of
$X\,B_{\perp}$ and $B_n$ projected along the line of sight. The
measured RM is then proportional to 
the sum of the ingress (far hemisphere) and egress (near hemisphere)
components. Equation~\ref{eqn:RM_1} becomes a function of polar
coordinate ($\phi$,\,$\zeta$), compression factor $X$, electron density
$n_e$, magnetic field strength $B_0$ and the angle of the magnetic
field to the plane of the sky $\Theta$ 
\small
\begin{equation}\label{eqn:RM_los}
  {\rm RM} = 0.81\,f\,\left(\frac{B_{\rm ||}(\phi,\zeta,{B_0}, \Theta,
    X)}{\muG}\right)\left(\frac{n_e}{\cmmthree}\right)\left(\frac{L(\phi,dr)}{pc}\right).
\end{equation}
\normalsize
The model implicitly assumes that the same $B_{\rm ||}$ applies everywhere
along the half-chord between the outer surface and mid-plane of
the shell (but different for ingress and egress). This assumption will
only be realistic for a thin shell; a more sophisticated
analysis is outside the scope of this paper. The model also assumes a
constant value for the electron density within the shell and, because
$n_e$, $B_{\rm los}$ and $f$ are degenerate in
Equation~\ref{eqn:RM_1}, the electron density and filling factor must
be estimated from independent data, if possible (see $\S\ref{sec:ne}$,
below).
 
Fig.~\ref{fig:bubble_grid_lrg} presents a grid of near-field
bubble models, illustrating how the RM observed on the sky changes as
$\Theta$ and $X$ are varied. In the far-field case, the
rotation-measure profile is spherically symmetric, i.e., constant with 
$\zeta$ (the angle between the profile sampling line and the Galactic
plane, see Fig.~\ref{fig:bubble_geom}). However, in the near-field
case the profile shape depends on both $\zeta$ and  $\Theta$ (the
orientation of the ordered magnetic field vector). As $\Theta$
increases from $0\degrees$ ($\overrightarrow{B_0}$ in the plane of the
sky) to $90\degrees$ ($\overrightarrow{B_0}$ pointing towards observer) the
large scale distribution of RMs changes 
from being anti-symmetric to symmetric around the central latitude
axis. Increasing $X$ leads to an overall increase in the magnitude of
the RM and a large difference in RM between the centre and inner-edge
of the shell. The shape of the profile edge also becomes more rounded
because of limb-brightening, although this would be much more
noticeable in far-field bubbles. 

In $\S\ref{sec:fitting}$, below, we present a maximum-likelihood
method used to fit the model to rotation measures from the
\citet{Taylor2009} catalogue. We incorporate estimates of $n_e$ from
$\halpha$ data as a prior, assuming Gaussian errors.

\subsection{$n_e$ from $\halpha$ data}\label{sec:ne}
From Equation~\ref{eqn:RM} we see that $n_e$ and $B_{||}$ are
degenerate and cannot be determined individually from observations of
RM alone. However, a separate observation of the emission measure
provides an independent estimate of the electron density along
the line of sight. EM is related to $n_e$ via
\begin{equation}\label{eqn:EM}
  {\rm EM} = \int_{0}^{\infty}n_e^2\,dl~~~{\rm pc\,cm^{-6}}.
\end{equation}
Assuming the same clumpy medium and geometry as presented in
Section~\ref{sec:RM_as_probe} this becomes
\begin{equation}\label{eqn:EM_1}
  {\rm EM} = n_e^2\,f\,L~~~{\rm pc\,cm^{-6}},
\end{equation}
with filling factor $f$ and path length $L$ as before.
EM may be calculated directly from the intensity of the
$\halpha\,(3-2)$ line via the equation of \citet{Reynolds1988}
\begin{equation}\label{eqn:em_halpaha}
  {\rm EM} = 2.75\left(\frac{T_e}{10^4\,{\rm K}}\right)^{0.9}\left(\frac{I_{\rm
  H_{\alpha}}\,K}{\rm R}\right)~~~{\rm cm^{-6}\,pc},
\end{equation}
where $T_e$ is the electron temperature in K, $I_{\rm H_{\alpha}}$ is
the intensity of the $\halpha$ emission in 
Rayleighs and
$K=\tau\,/\,(1-e^{-\tau})$ is a correction term to account for dust
extinction between the $\halpha$ emission and the observer (see
Section~\ref{sec:ha_extinction}).

Most estimates of $T_e$ for the Gum Nebula within the literature vary
between 6500\,K and 11500\,K. Electron temperatures derived  by
\citet{Reynolds1976a} from a comparison of $\halpha$ to [NII]
linewidths are consistent with a uniform temperature of 11300\,K
throughout the nebula. However, \citet{Vidal1979} determined the
electron-temperature to be $T_e=6500$\,K from existing optical
emission-line data. We adopt a uniform value of $8000$\,K for this
analysis.

\begin{figure*}
  \centering
  \includegraphics[width=8.8cm, trim=0 0 0 0]{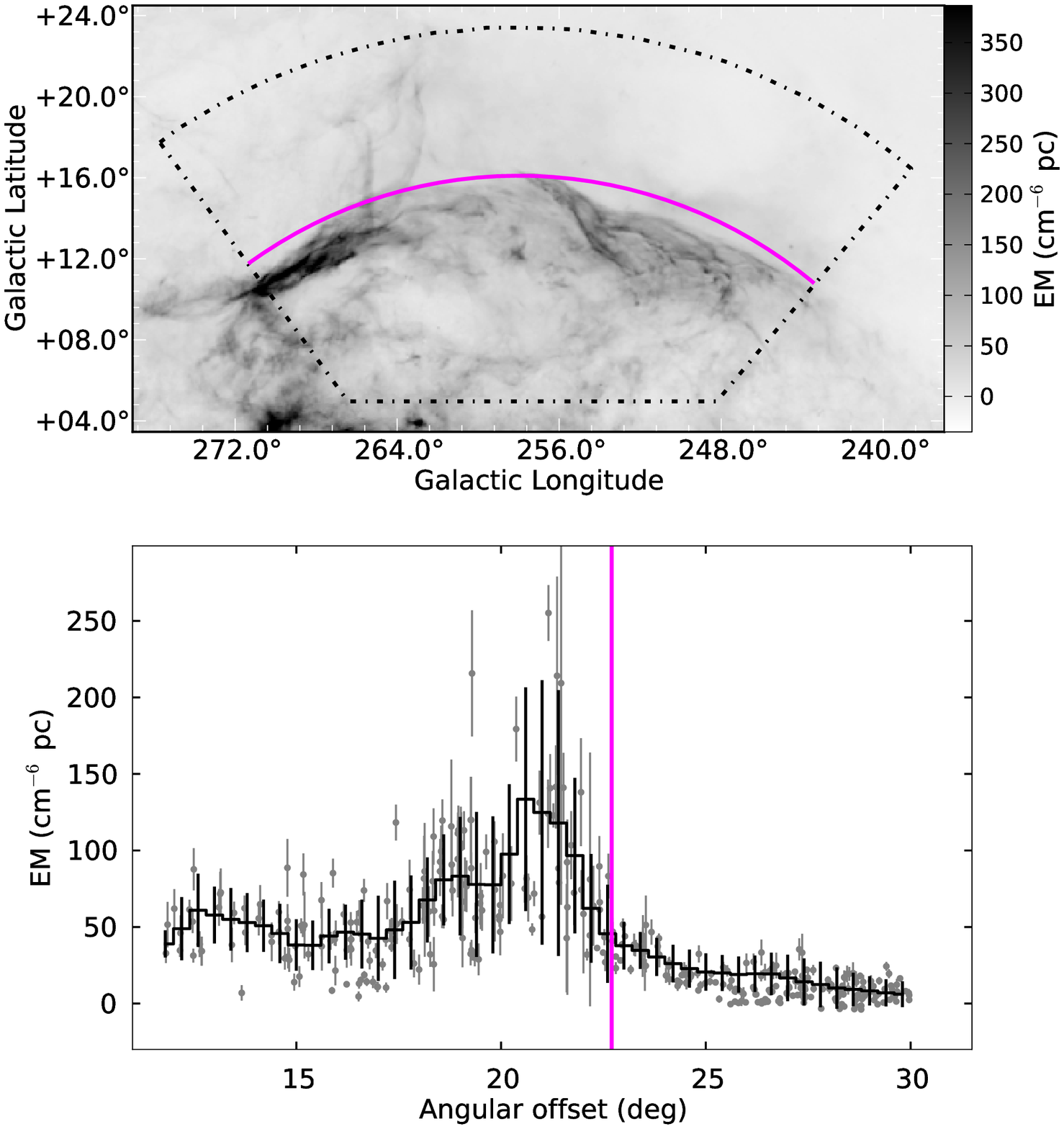}
  \includegraphics[width=8.8cm, trim=0 0 0 0]{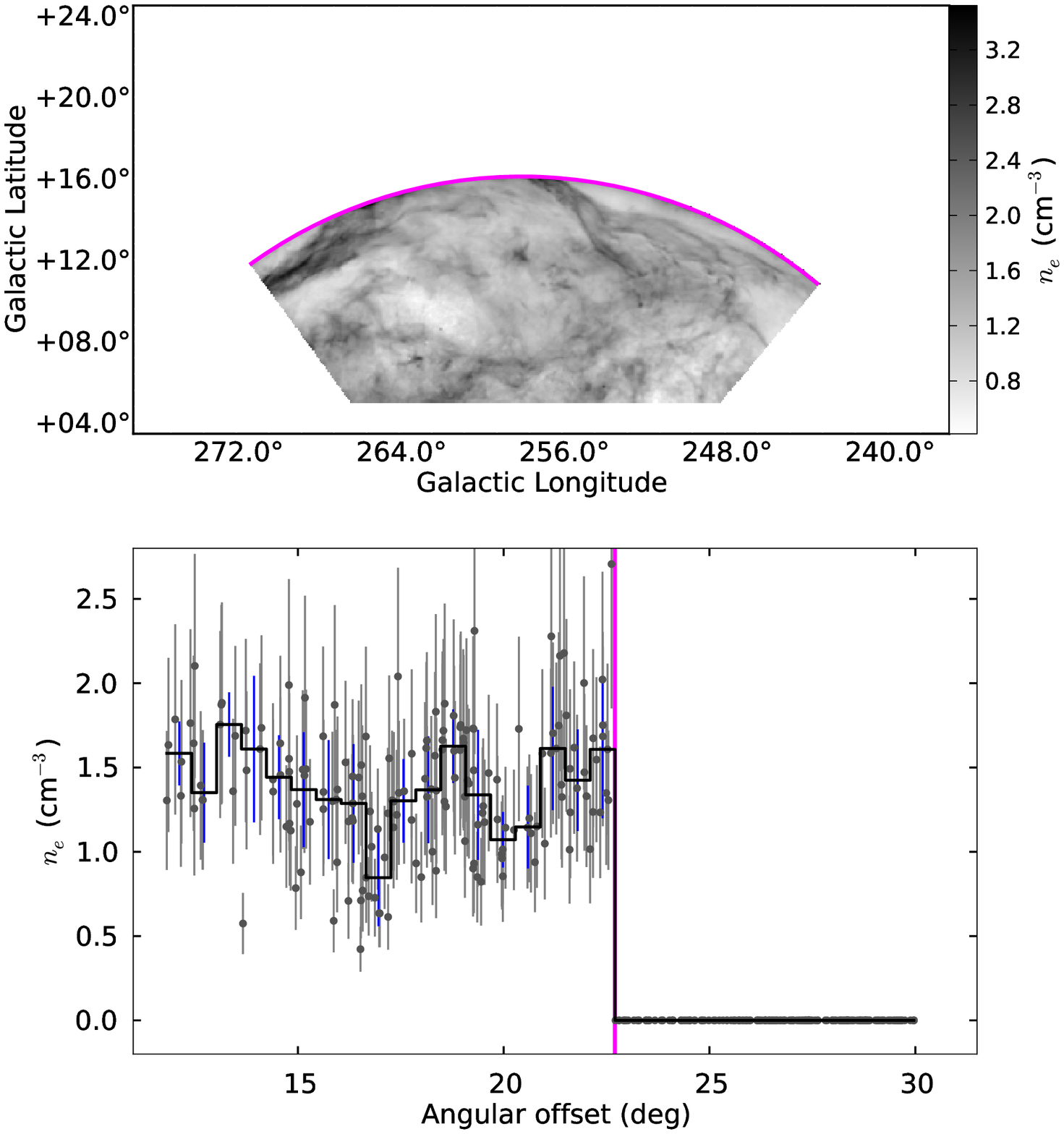}
  \caption{\small {\it Left}: EM map and azimuthally averaged
    profile of the upper Gum region. The individual grey points
    in the bottom panel show the EM values at the positions of the RM
    samples while the black points with error bars show the azimuthal
    average and standard-deviation in 0.2 degree bins. {\it Right}:
    Electron density map and azimuthal averaged profile derived
    assuming the best-fit parameters in Table~\ref{tab:results}. The
    magenta line shows the edge of the shell model.}
  \label{fig:EM_selected}
\end{figure*}
Fig.~\ref{fig:EM_selected}\,-\,{\it top-left} presents the emission
measure map of the upper Gum region derived from the
\citet{Finkbeiner2003} $\halpha$ data
using Equation~\ref{eqn:em_halpaha}. Azimuthally averaged EM values 
peak at $220\,\pccmmsix$ (see Fig.~\ref{fig:EM_selected}\,-\,{\it
  bottom-left}), falling to $\sim80\,\pccmmsix$ in the interior and
$\lesssim30\,\pccmmsix$ outside the nebula. Our values are largely consistent
with those of previous authors. \citet{Reynolds1976a} measured the EM
via pointed $\halpha$ spectral observation and found it varied from
$100\,\pccmmsix$ interior to the nebula to $240\,\pccmmsix$ in the
upper arc. Scans across the region by \citet{Reynolds1976a} at
$l=240\degrees$ determined the Galactic background to be
$<10\,\pccmmsix$ rising to $28\,\pccmmsix$ at the mid-plane.

Assuming the geometry $L(\phi,dr)$ from the best-fit to our shell
model (presented in $\S\ref{sec:results}$, below) and best-fit filling
factor $f=0.3$ we use Equations~\ref{eqn:EM_1} and~\ref{eqn:path_chord} to 
derive the electron density $n_e$ inside the clumpy shell. The
electron density map and azimuthally averaged profile are presented in
Fig.~\ref{fig:EM_selected}\,-\,{\it right}. The fitting
procedure takes the value of $n_e$ as a prior (see
$\S\ref{sec:fitting}$) altering the most likely shell
geometry and filling factor $f$, requiring a new estimate of $n_e$. To
correct for this inconsistency we re-calculated $n_e$ using the new
$L(\phi,dr)$ and $f$, and iterated over the fitting loop until all
values converged. The final electron density 
was determined to be $n_e=1.4\pm0.4\,\cmmthree$, which compares well
with \citet{Reynolds1976b} who found $n_e\approx1\,\cmmthree$
in the fainter parts of the nebula, assuming fixed 
physical parameters ($\phi_{outer}=18\degrees$, $D=450$\,pc and
$15\lesssim dr\lesssim30\,$pc).

\subsection{Maximum likelihood analysis}\label{sec:fitting}
We use a Markov Chain Monte Carlo (MCMC) algorithm to fit the model
shell to the RM data in the upper Gum region. The de-facto algorithm
for performing MCMC fitting is the Metropolis-Hastings
algorithm \citep{Metropolis1953,Hastings1970}, which randomly samples
over parameter space, accepting or rejecting models based on their
likelihood $\mathcal{L}$ (i.e., the 
probability of the data given the model parameters). New positions
with greater $\mathcal{L}$ than previously are always accepted, while
those with smaller $\mathcal{L}$ are occasionally accepted. Our code
makes use of the efficient affine-invariant sampler \citep{Goodman2010}
implemented in the {\scriptsize
  EMCEE}\footnote{http://dan.iel.fm/emcee/} {\it python} module by  
\citet{Foreman-Mackey2013}. {\scriptsize EMCEE} controls a number of
parallel samplers, referred to as `walkers', each of which corresponds
to a vector of free parameters within the model. The walkers are
initialised to a point in n-dimensional parameter space and are 
iteratively updated to map out the probability distribution. At each
iteration the likelihood is calculated assuming Gaussian errors
according to $\mathcal{L}=e^{-\chi^2/2}$, where $\chi^2$ is the
standard chi-squared goodness-of-fit statistic. If  priors with
measured uncertainties exist for any model parameter, we incorporate
them into the likelihood calculation by summing their chi-squared values
$\chi^2=\chi^2_{\rm model}+{\textstyle\sum\nolimits_{i}}\,\chi^2_{{\rm prior}\,i}$.

The fitter is started by generating 300 walkers initialised to random
values of the free parameters. The MCMC code is initially run for 400
`burn-in' iterations to allow the walkers to settle in a clump around
the peak in likelihood space. The fitting routine is then run 
for 10000 iterations to produce a well-sampled likelihood
distribution. We determine the best fitting model from the mean of the 
marginalised posterior distribution for each free parameter. The
$\pm1\sigma$ uncertainties are calculated as the
fractional positions at $1-{\rm erf}(1\sigma)=0.1572$ and
${\rm erf}(1\sigma)=0.8427$ on the normalised cumulative
distribution. Our results are presented in $\S\ref{sec:results_fit}$

\subsubsection{Scatter in RM as a hyperparameter}
The median measurement uncertainty on the selected RM data is
$\sigma({\rm RM})=12\,\rmtwo$, considerably smaller than the scatter
evident in Fig.~\ref{fig:RM_selected}\,-\,{\it right}. The error-bars
reflect only the uncertainty in the measurement and do not take into
account systematic scatter, e.g., due to fluctuations in $B_{||}$ or
$n_e$ on scales much smaller than the sampling grid, or systematic
errors in RM determination. We characterise
this additional variation using a term $\delta({\rm RM})$ added in
quadrature to the measurement uncertainty. This new scatter term is
included in the model as a free parameter, however, it is treated
slightly differently when calculating the likelihood function. 

\citet{Lahav2000} and \citet{Hobson2002} present a formalism for
performing joint analysis of cosmological datasets by introducing 
{\it hyperparameter} weighting terms, the values of which are
determined directly from the statistical properties of the data. This
approach is easily adapted to find the self-consistent uncertainties
for data with ill-determined error-bars. Using Equations~29 and~30 of
\citet{Hobson2002} we calculate a likelihood function using
the modified chi-squared statistic
\begin{equation}
  {\rm \chi^2 = \sum_i\left[\,\frac{(RM_i-RM_{mod})^2}{\sigma(RM)_{tot,i}^2} 
  + {\rm ln}(2\,\pi\,\sigma(RM)_{tot,i}^2)\,\right]},
\end{equation}
where ${\rm RM_i-RM_{mod}}$ is the difference between the ${\rm
  i^{th}}$ rotation measure and the model at that position. The second
term inside the parentheses is required to correctly normalise the
likelihood and the total uncertainty on the ${\rm RM_i}$ is given by
\begin{equation}
  {\rm \sigma(RM)_{tot,i}^2=\sigma(RM)_i^2+exp[\,2\,ln(\,\delta(RM)_i\,)\,]}.
\end{equation}
Here we solve for ${\rm ln(\,\delta(RM)\,)}$ rather than directly for
${\rm \delta(RM)}$ so as to enforce positivity in the scatter term
(i.e., uncertainties cannot be negative).

\section{Results}\label{sec:results}

\subsection{General comparison of model and data}
Before presenting the results of the MCMC analysis, it is useful to
visually compare the RM data shown in Fig.~\ref{fig:RM_selected}
with the simple shell models illustrated in
Fig.~\ref{fig:bubble_grid_lrg}. Two
key discriminators stand out in the behaviour of the models. Firstly,
the difference in rotation measure between the interior and
the peak of the shell (${\rm \Delta RM}$, illustrated by the black line in
Figures~\ref{fig:bubble_grid_lrg}) is a strong function of the
compression factor $X$, assuming other 
parameters are fixed. From Fig.~\ref{fig:RM_selected}\,-\,{\it right} we
see that the measured ${\rm \Delta RM}$ in the profile of the northern Gum
Nebula is $\sim350\,\rmtwo$, restricting the compression factor to
$X\lesssim4$. Models with higher values of $X$ result in much greater
${\rm \Delta RM}$ for all reasonable values of $n_e$, $f$, $B_0$ and
$\Theta$. Secondly, the longitudinal RM-gradient is directly related
to the pitch angle of the magnetic field. This behaviour is due to the
close proximity of the nebula, so that sight-lines from opposite sides
are not parallel and intersect a uniform field at different angles.
If the ordered magnetic field is directed along the plane of the sky
at the nebula's centre ($\Theta=0\degrees$), we would expect to
measure equal positive and 
negative RMs on either side of the central longitude. For a magnetic
field pointing directly towards ($\Theta=90\degrees$) or away
($\Theta=-90\degrees$) from the Sun the RMs would display symmetric
positive or negative patterns, respectively. We see mostly positive
RMs towards the Gum Nebula, with a slight positive gradient towards
lower Galactic longitudes (see Fig.~\ref{fig:RM_selected}). From an
examination of the grid of models shown in
Figure~\ref{fig:bubble_grid_lrg}, we can conservatively state that the
ordered magnetic field is pointing towards the Sun at an angle
$\Theta\gtrsim20\degrees$. This is because the negative peak in RMs at
positive Galactic longitudes is absent from models with
$\Theta\gtrsim20\degrees$, and from the \citet{Taylor2009} RMs. In
$\S\ref{sec:results_fit}$ below we quantify these assertions using
fits to the data.

\subsection{Fits to the model shell}\label{sec:results_fit}
Here we present the results of fitting the model described in 
$\S\ref{sec:mag_bubble}$ to a subset of the \citet{Taylor2009} RM
catalogue. The RM data included in the fit are outlined by the
wedge-shaped box in Fig.~\ref{fig:RM_selected}\,-\,{\it left}
and was selected to bracket the nebula above
$b>5\degrees$ (excluding the `stalk' region and a handful of negative
outliers - see $\S\ref{sec:RMs_selected}$). We fixed the centre of the
model to $(l,\,b)=(258.0\degrees,\,-6.6\degrees)$ so that the
circumference of  the shell corresponds to the sharp outer edge seen
in the $\halpha$ data (see Fig.~\ref{fig:gum_summary_halpha}). The
mean of the marginalised likelihood distribution is a good estimator
of the best fitting value for each parameter; these are reported
in Table~\ref{tab:results} and described below.

\begin{table*}
  \caption{Results of fitting the ionised shell model to the RM catalogue.}
  \label{tab:results}
  \begin{small}
  \begin{tabular}{lccc c@{~~}r@{\,}l c@{~~~~}r@{\,}l c@{~~~~~~~~}r@{\,}l}
    \tableline
    \tableline
    \rule[-0.3em]{0pt}{1.35em}(1)       & (2)    & (3)  & (4) & \multicolumn{3}{c}{(5)} & \multicolumn{3}{c}{(6)} & \multicolumn{3}{c}{(7)}\\
    \rule[-0.5em]{0pt}{1.55em}Parameter & Symbol & Unit & Notes &\multicolumn{9}{c}{Assumed Background Level} \\
    \cline{5-13}
    \rule[-0.5em]{0pt}{1.75em} & & & &\multicolumn{3}{c}{Flat} &\multicolumn{3}{c}{\citet{Sun2008}} &\multicolumn{3}{c}{\citet{Jansson2012}} \\
    \tableline
    \rule[0mm]{0pt}{1.1em}Distance  & $D$  &  pc         &  Fixed &&  450 & &&  450 & &&  450\\
    \rule[-0.6em]{0pt}{1.75em}Background RM   & ${\rm RM_{bg}}$ & \rmtwo          &  Fixed && $-26.4$ & && $-64.0$ & && $-36.6$\\
    \rule[-0.6em]{0pt}{1.75em}Filling factor  & $f$              & --             & Free && 0.4 & $^{+0.3}_{-0.2}$  && 0.3 & $^{+0.3}_{-0.1}$ && 0.2 & $^{+0.2}_{-0.1}$ \\
    \rule[-0.6em]{0pt}{1.75em}Angular radius  & $\phi_{\rm outer}$ & deg.          & Free && 22.7 & $^{+0.2}_{-0.1}$ && 22.7 & $^{+0.1}_{-0.1}$ && 22.7 & $^{+0.1}_{-0.1}$\\
    \rule[-0.6em]{0pt}{1.75em}Shell thickness & $dr$             & pc             & Free && 20.2 & $^{+1.8}_{ -1.6}$ && 18.5 & $^{+1.5}_{ -1.4}$ && 18.5 & $^{+1.3}_{ -1.3}$ \\
    \rule[-0.6em]{0pt}{1.75em}Field angle     & $\Theta$         & deg.           & Free && 55 & $^{+15}_{-12}$ && 43 & $^{+13}_{-9}$ && 55 & $^{+16}_{-12}$ \\
    \rule[-0.6em]{0pt}{1.75em}Field strength  & $B_0$          & $\mu$G         & Free && 8.8 & $^{+6.1}_{-4.0}$ && 3.9 & $^{+4.9}_{-2.2}$ && 3.9 & $^{+4.2}_{-2.1}$ \\
    \rule[-0.6em]{0pt}{1.75em}Compression factor & $X$           & --             & Free && 1.1 & $^{+0.5}_{-0.3}$ && 6.0 & $^{+5.1}_{-2.5}$ && 6.8 & $^{+5.3}_{-2.8}$ \\
    \rule[-0.6em]{0pt}{1.75em}Electron density & $n_e$           & ${\rm cm^{-3}}$ & Prior && 1.4 & $^{+0.4}_{-0.4}$ && 1.3 & $^{+0.4}_{-0.4}$ && 1.2 & $^{+0.4}_{-0.4}$ \\
    \rule[-0.6em]{0pt}{1.75em}Additional RM Scatter & $\delta({\rm RM})$ & \rmtwo & Free && 75.1 & $^{+2.9}_{-2.7}$ && 71.0 & $^{+2.7}_{-2.7}$ && 70.0 & $^{+2.9}_{-2.6}$ \\
    \tableline
  \end{tabular}
\end{small}
\end{table*}

\begin{figure*}
  \centering
  \includegraphics[width=18.5cm, angle=0, trim=0 0 0 0]{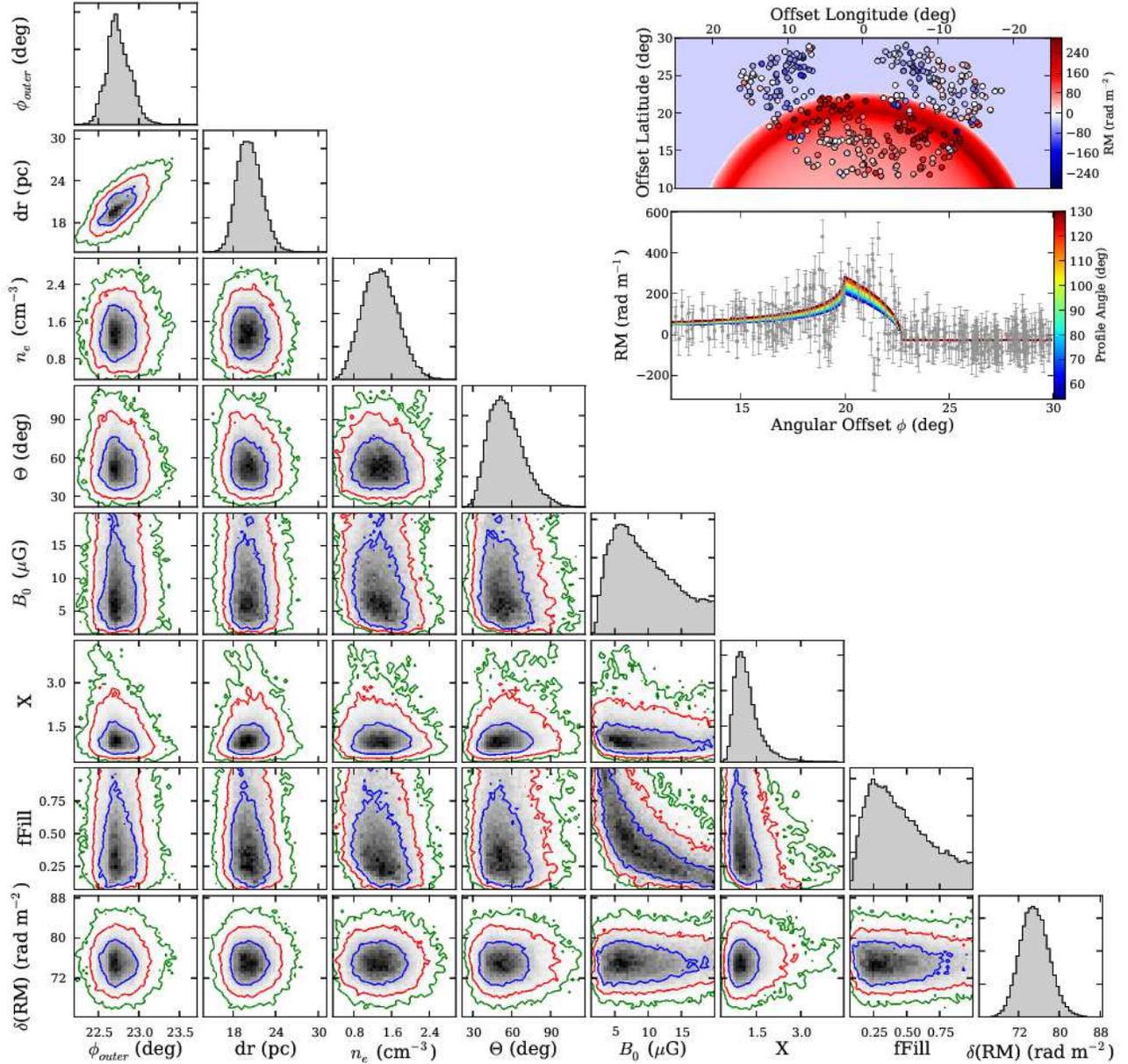}
  \caption{\small {\it Main-figure:} Triangular grid of confidence
    plots for the Gum Nebula, produced by fitting the ionised shell
    model to the \citet{Taylor2009} RM-catalogue assuming a flat
    background (also see Table~\ref{tab:results}). The blue, red and green 
    contours represent the $1\sigma$, $2\sigma$ and $3\sigma$
    confidence intervals, respectively. The filled histograms on the
    diagonal show the marginalised likelihood distributions for each
    free parameter in the model. {\it Inset:} The rotation measure
    image ({\it top}) and azimuthal profile ({\it bottom}) for the
    best fitting model over-plotted by the selected RM data from the
    \citet{Taylor2009} catalogue. Note the compressed range in angular
  offset on the x-axis of the inset panel, compared to the equivalent
  plots in Fig.~\ref{fig:bubble_grid_lrg}.}
  \label{fig:triangle_flat}
\end{figure*}

We initially ran our MCMC fitting procedure assuming a flat,
large-scale background of ${\rm RM_{bg}=-26.4\,\rmtwo}$ and with all
other parameters free, except distance, which was fixed at $D=450$\,pc
and electron density, for which a prior of
$n_e=1.4\pm{0.4}\,\cmmthree$ was set (see $\S\ref{sec:ne}$). Likelihood
distributions $\mathcal{L}$ and plots of RM are presented in
Fig.~\ref{fig:triangle_flat}. The triangular matrix of confidence
contour plots illustrates how the free parameters 
interrelate, while the histograms on the diagonal show the 
marginalised likelihood distributions for individual parameters. Of
particular note are the distributions for $f$, $n_e$, $B_0$, $dr$ and
$X$. The likelihood distribution for the filling factor is very broad,
only constraining $f\gtrsim0.25$, below which the marginalised
distribution drops off rapidly. The curved and elongated confidence
contours between $B_0$ and $f$ mean that these two parameters are
highly degenerate, and that the strength of the magnetic field is
not well determined in the absence of an independent estimate of
$f$. The prior on electron density has the effect of constraining the
likely range of $n_e$ values and eliminates most of the degeneracy between
$n_e$ and $B_0$. Confidence contours between $dr$ and $\theta_{\rm
  outer}$ are elliptical in shape, indicating that the RM-data do
not pinpoint the radius and thickness of the shell
independently. Nonetheless, the range of values for each parameter is
small and their absolute values are well constrained. The marginalised
$\mathcal{L}$ distribution for the compression factor $X$ exhibits a
narrow profile centred on 1.1, implying that the gas within the shell
has not been significantly compressed. The hyperparameter
characterising the additional scatter on the RMs is
well determined at $\delta({\rm RM})=75.1^{+2.9}_{-2.7}\,\rmtwo$, as
shown by the Gaussian form of its marginalised likelihood
distribution. By definition the hyperparameter procedure adjusts
$\delta({\rm RM})$ so that $\chi^2=1.0$, thus $\delta({\rm RM})$ can
be though of as a proxy for $\chi^2$ when comparing `goodness-of-fit'
between models.

\begin{figure*}
  \centering
  \includegraphics[width=18.0cm, angle=0, trim=0 0 0 0]{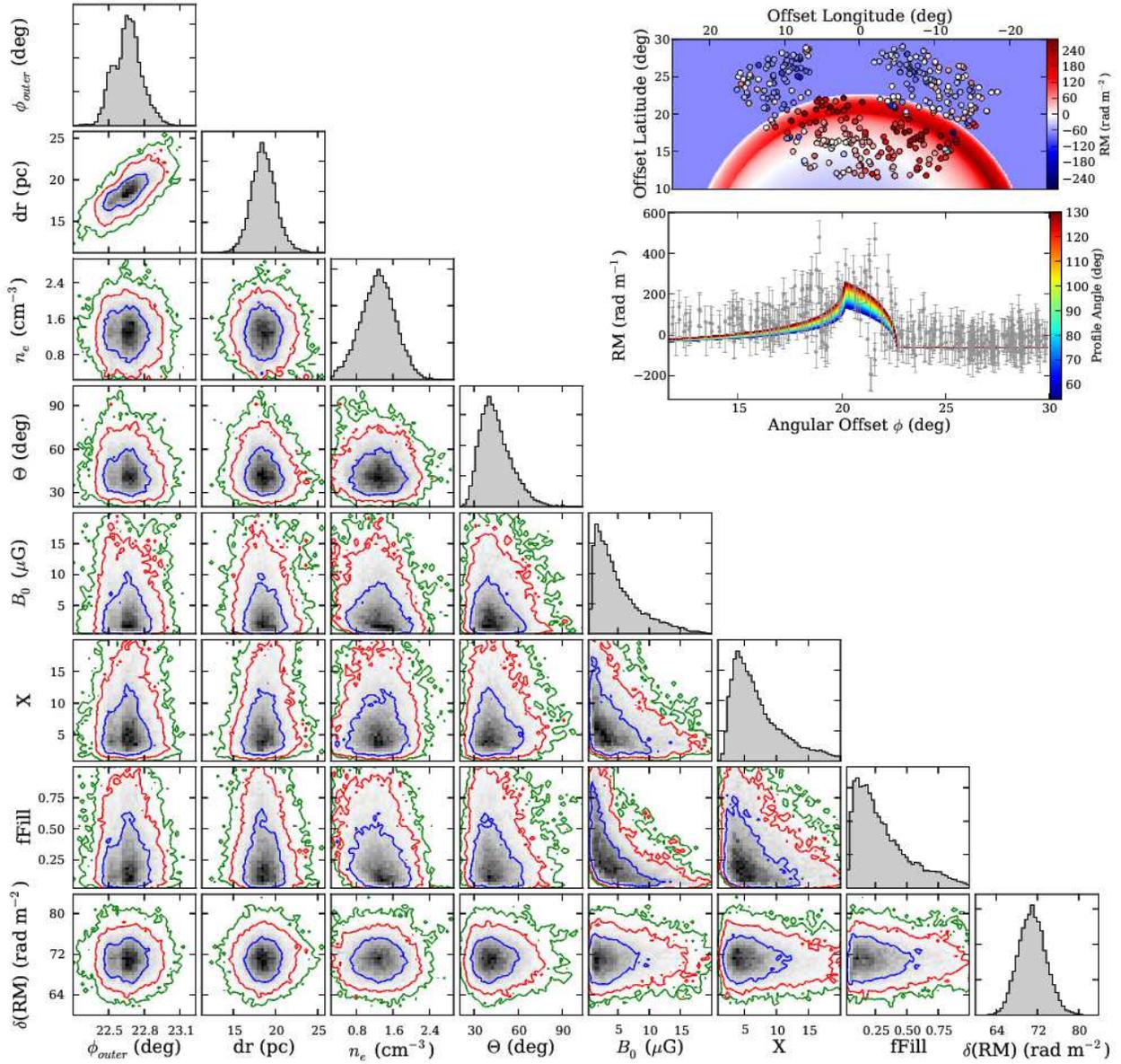}
  \caption{\small As for Fig.~\ref{fig:triangle_flat}, but assuming a
    background as given by \citet{Sun2008}.}
  \label{fig:triangle_Sun}
\end{figure*}

\begin{figure*}
  \centering
  \includegraphics[width=18.0cm, angle=0, trim=0 0 0 0]{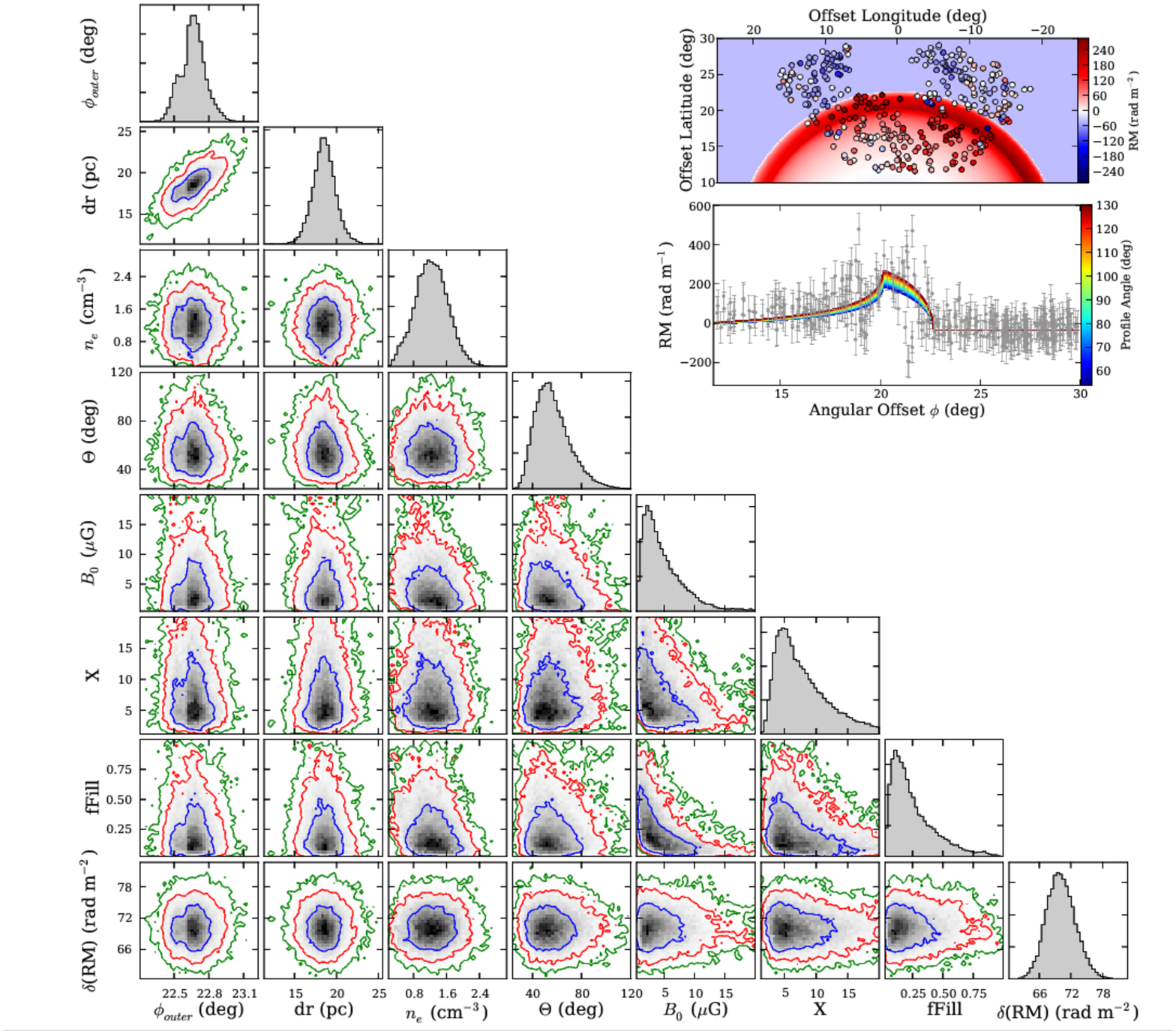}
  \caption{\small As for Fig.~\ref{fig:triangle_flat}, but assuming a
    background as given by \citet{Jansson2012}.}
  \label{fig:triangle_Jansson}
\end{figure*}

We ran the MCMC analysis again after correcting the data for the
large-scale RM-gradients modelled by \citet{Sun2008} and 
\citet{Jansson2012}. The shape of the gradients is similar for both
modes (illustrated in Fig.~\ref{fig:RM_profile_bgs}) and
removing these backgrounds has the effect of decreasing
the RM-values towards lower Galactic latitudes. Due to the
orientation of the selection box, this results in a 
decreased RM-signal towards the centre of the Gum Nebula compared to
the edge. The parameters of the best-fitting models to the
gradient-subtracted versions of the RM-catalogue are presented in
columns (6) and (7) of Table~\ref{tab:results}, and illustrated in
Figs.~\ref{fig:triangle_Sun} and~\ref{fig:triangle_Jansson}. The
results of both MCMC fits are identical within the errors, so we
refer to the version assuming the \citet{Sun2008} background in
the following discussion.

Comparing the results of fits to the flat- and gradient-subtracted
data, the most significant difference is in the compression factor
$X$, whose value changes from $1.1^{+0.5}_{-0.3}$ to $6.0^{+5.1}_{-2.5}$,
respectively. This change is purely a result of the smaller difference
between RMs in the interior of the nebula and the peak. $X$ is much
less well constrained by the gradient-subtracted data as the
difference between the interior and exterior levels approaches the
scatter on the data. The confidence 
contours in Fig.~\ref{fig:triangle_Sun} also show that $X$ is more
degenerate with $f$ and $B_0$. The higher compression factor is
balanced by corresponding small decreases in filling factor $f$, shell
thickness $dr$, field strength $B_0$ and electron density $n_e$. The
filling factor $f$ is slightly better constrained, leading to a
correspondingly more precise value for
$B_0=3.9^{+4.9}_{-2.2}\,\muG$. The uncertainties on fitted values of
magnetic field angle $\Theta$ 
are large (typically $\sim12\degrees$), however, the fitted angles are
broadly similar for all three fits. The value of $\delta({\rm RM})$ is
lower by $4.1\,\rmtwo$ in the gradient-subtracted fit, implying that the
model is a better match to this data. However, the absolute change is
equivalent to only $\sim1.5\sigma$ between the two MCMC runs. We
discuss the implications of the results in $\S\ref{sec:discussion}$.

\subsubsection{Consistency checks - pulsars}\label{sec:pulsar_chk}
Although fewer in number than extragalactic sources, pulsars with well
determined distances are useful in checking the results of an
extragalactic RM or EM analysis. The interaction between free
electrons and photons introduces a differential time delay $\Delta t$
across the observational bandwidth $\Delta\nu$. The delay is a
function of frequency $\nu$ and is characterised by the dispersion
measure according to $(\Delta t/\mu s)=8.3\,\Delta \nu\,({\rm 
 DM}/cm^{-3}\,pc)\,\nu^{-3}$, where all frequencies are in GHz. The
measured DM of a pulsar is related to the electron density via
\begin{equation}\label{eqn:DM}
  {\rm DM} = \int_{obs}^{psr}n_e\,dl~~~{\rm cm^{-3}\,pc}.
\end{equation}
Assuming the same volume filling factor $f$ and path length $L$ as
before, Equation~\ref{eqn:DM} can be written as
\begin{equation}\label{eqn:DM_1}
  {\rm DM} = n_e\,f\,L~~~{\rm cm^{-3}\,pc}.
\end{equation}
With suitable observations and by combining Equations~\ref{eqn:RM_1},
\ref{eqn:EM_1} and \ref{eqn:DM_1} we can solve for $f$,
$B_{||}$ or $n_e$ along the line-of-sight to a pulsar. For example,
the average line-of-sight magnetic field strength is given by
$B_{||}=\frac{\rm RM}{\rm 0.81\,DM}$, the electron density inside the clumps
by $n_e=\frac{\rm EM}{\rm DM}$ and the filling factor by
$f=\frac{\rm DM^2}{{\rm EM}\,L}$.

The RMs, DMs and accurate distances (where available) of selected
pulsars towards the upper Gum region have been presented in
Fig.~\ref{fig:pulsar_map}. A few general trends are worth noting: on
average the DMs increase towards the Galactic mid-plane as the
electron density peaks at $b\approx0\degrees$
\citep{Gaensler2008}. After taking the latitude dependence into
account, the DMs of pulsars towards the nebula appear to
be enhanced when compared to sight-lines outside of its
border. Typically, pulsar sight-lines just inside shell edge have DMs
of 100\,--\,150\,$\pccmmthree$ compared to 20\,--\,60\,$\pccmmthree$
outside. The DM of the Vela pulsar (J0835$-$4510) is
68\,$\pccmmthree$, greater by 19\,$\pccmmthree$ compared to the pulsar
J0737$-$3039A, which lies just outside the nebula 
at distance of 1.1\,kpc. The difference is representative
of how much dispersion is created by one wall of the
nebula.

The clumpy electron density derived from
$n_e=\frac{\rm EM}{\rm DM}$ using only pulsars above $b>5\degrees$ inside the
nebula varies from $0.7\,\cmmthree$ to $2.2\,\cmmthree$ with an average of
$1.5\,\cmmthree$. Outside the nebula, but away from the plane, $n_0$
(i.e., the ambient value) falls to values of $0.2\,\cmmthree$ to
$0.9\,\cmmthree$. Given the large uncertainties these values are in
agreement with our best-fitting models.

Only a handful of pulsars inside the nebula have both RMs and DMs,
allowing the determination of the average line-of-sight magnetic field
strength $B_{||}$. Three pulsar sight-lines intersect the upper Gum region,
away from the shell, and their DM values suggest they lie beyond the
nebula. Values for $B_{||}$ derived from the pulsars range between
$0.9\,\muG$ and $2.4\,\muG$, significantly lower than the best-fit
value to the flat-background data ($8.8^{+6.1}_{-4.0}\,\muG$), but
consistent with the value found when fitting the gradient-subtracted
RM-data ($3.9^{+4.9}_{-2.2}\,\muG$).  Some of the discrepancy may be
explained by our choice of filling factor, which is not
well-constrained in any of the results. The fitted value of $B_0$ is
highly dependent on $f$ and a values of $f\approx0.5$ would bring
our models into agreement with $B_{||}$ derived from
pulsars. Determining the filling factor from a pulsar requires a good
estimate of the path length and hence of the distance to the
pulsar. Unfortunately, no pulsars with well measured distances lie
towards northern part of the Gum Nebula, thus we do not attempt to
estimate $f$ at this time.

In summary, we find that the pulsar data are consistent with our
results, and especially favour the datasets which have had a
model large-scale Galactic RM signature subtracted (i.e., columns 6
and 7 in Table~\ref{tab:results}, fits in
Figures~\ref{fig:triangle_Sun} and~\ref{fig:triangle_Jansson}).

\section{Discussion and further analysis}\label{sec:discussion}
The best-fitting shell models have some interesting implications,
especially for the local direction of the ordered magnetic field and
the fitted compression factor. We 
discuss the results below, but start by noting the limitations of the
model and the data. We further analyse the results by comparing the
expected radio-continuum signature of the best-fitting models to the
diffuse S-PASS 2.3\,GHz data.

\subsection{Limitations of the simple shell model}\label{sec:model_limitations}
The simple ionised shell model presented here has a number of
limitations and assumptions that should be considered when
interpreting the results. We have already noted in
$\S\ref{sec:results_fit}$ that the compression factor $X$ is sensitive to
differences in RM between the rim, exterior and interior of the
nebula. The vertical orientation of 
the selection box and the relatively narrow portion of the nebula
sampled by observations (a $\sim76\degrees$ pie-shaped sector, see
Fig.~\ref{fig:RM_selected}\,-\,{\it left}) mean that latitudinal gradients
in RM affect the fitted value of $X$ in 
particular. Similarly, the fitted angle of the 
magnetic field $\Theta$ is directly dependent upon the observed
longitudinal RM-gradient, as the ordered field runs parallel to
the Galactic disk \citep{Mathewson1970,Han2006}. Thus, isolating the RM-signal of
the Gum Nebula is a critical step in our analysis and involves
subtracting RMs due the large-scale Galactic background, and the
smaller-scale magneto-ionic material along the line of sight. A
residual RM-gradient remaining within the data would skew the values
of $X$ and $\Theta$ derived from our MCMC analysis.

The lack of coverage below ${\rm Dec.=-40\degrees}$ in the
\citet{Taylor2009} RM catalogue makes determining an accurate
large-scale Galactic background difficult, so we have tried two approaches:
subtracting a flat background and subtracting a model
RM-gradient. Exterior to the Gum Nebula ($b\gtrsim12\degrees$) the
RM-data are small and negative, consistent with a homogeneous
background. However, the RM values must increase towards the
Galactic mid-plane, as the electron density is known to fall
exponentially with increasing latitude
\citep{Cordes2002,Gaensler2008}. As discussed in
$\S\ref{sec:RMs_selected}$ previously, the Galactic RM-models of
\citet{Sun2008} and \citet{Jansson2012} represent the best existing
estimate of the RM distribution due to the bulk of the Galaxy behind
the nebula. Neither model is a good fit to the local RM
distribution, poorly matching RM-structures on scales of
$\sim10\degrees$ in the vicinity of the Gum Nebula. However, it is
encouraging that both the \citet{Sun2008} and \citet{Jansson2012}
models have similar gradients so we believe the large-scale morphology
to be reliable, but not the local calibration. Therefore, towards the
mid-plane, RMs with the flat background subtracted constitute an upper
limit on signal from the Gum Nebula. 

On small scales the division of RMs into `background', `Gum Nebula'
and `other object' categories is necessary to obtain a clear
RM-signal (see $\S\ref{sec:RMs_selected}$ and
Fig.~\ref{fig:RM_selected}). This identification procedure draws on all of
the available data to make informed decisions, but the process is
still somewhat subjective. Residual RMs from unidentified discrete
objects may still be present in the data, or the identified objects
may extend behind the footprint of the Gum Nebula. For example, the RM
signature of a small $\hii$~region ($\lesssim2\degrees$) overlaid on
the rim could be erroneously fitted as a gradient in $l$ or $b$, leading
to a systematic errors in $\Theta$ or $X$, respectively. Future
surveys that deliver a more accurate and densely sampled grid of RMs
covering the southern sky are required to resolve remaining ambiguities.

It is clear from the filaments visible in the $\halpha$ map that $n_e$
is structured on scales down to the $6'$ resolution of the image. This
clumpy distribution of electrons is accounted for in the model using a
global volume filling factor $f$, leading to an occupation length
$f\,L$ for all sight-lines. Variations in $n_e$ on scales much smaller
than the beam lead to fluctuations in RM, which manifest as an
additional uncertainty on the RMs, codified as $\delta({\rm RM})$ in
the model. Because the spatial sampling is coarse ($\sim1/{\rm
  degree}^2$), the value for $\delta({\rm
  RM})=75.1^{+2.9}_{-2.7}\,\rmtwo$ is an upper limit on the true
scatter in rotation measure. The high value may also reflect genuine
scatter of ISM properties between the model and the data.

We have assumed the electron density profile $n_e(r)$ within the shell is
constant as a function of radius. This is in line
with the description of a wind-blown bubble (see \citealt{Weaver1977},
Fig.~3), or with the physics of an expanding ionisation front in an
evolved $\hii$~region \citep{Draine2011}. However, a constant $n_e(r)$ is inconsistent
with the density profile of a Sedov-phase SNR, which increases
from the centre towards the shock-front
\citep{vanderSwaluw2001}. The measured 
density profile of the Gum Nebula presented in Fig.~\ref{fig:EM_selected}
agrees with a constant $n_e(r)$ within the errors. The density is
slightly enhanced towards the front edge of the shell, but only at a
$\sim1\sigma$ level, so we consider this assumption reasonable.

The model accounts for compression at the edge of the shell using a
factor $X$ by which both the ambient density $n_0$ and the component
of the magnetic field tangent to the shell surface ($B_{\perp}$) are
amplified. The model assumes that the gas inside the shell is 100
percent ionised by the powering source or the passing shock. The
magnetic field component that produces the RM signature ($B_{||}$) is
given by the projection of $B_{\perp}$ and the radial component,
$B_n$, onto the  line-of-sight. As a computational convenience
$B_{||}$ is assumed to have a constant strength throughout 
the thickness of each shell wall; an assumption which is valid only
for a thin shell. For the Gum Nebula, the ratio $R/dr\approx15$ and the
best-fit compression factor is low ($X<10$), so this assumption is
acceptable.

The model has implicitly assumed that the lines of the ordered Galactic magnetic
field are parallel to each other and to the disk of the Galaxy. If they
loop, converge or diverge significantly within the nebula ($260\,$pc)
then a much more sophisticated treatment is required, coupled with a
more finely-sampled grid of rotation measures. Such an analysis is
beyond the scope of this paper but should be considered with future
datasets.

\subsection{Orientation and strength of the ordered Galactic magnetic
  field}\label{sec:strB}
\begin{figure}
  \centering
  \includegraphics[width=8.5cm, angle=0, trim=0 0 0 0]{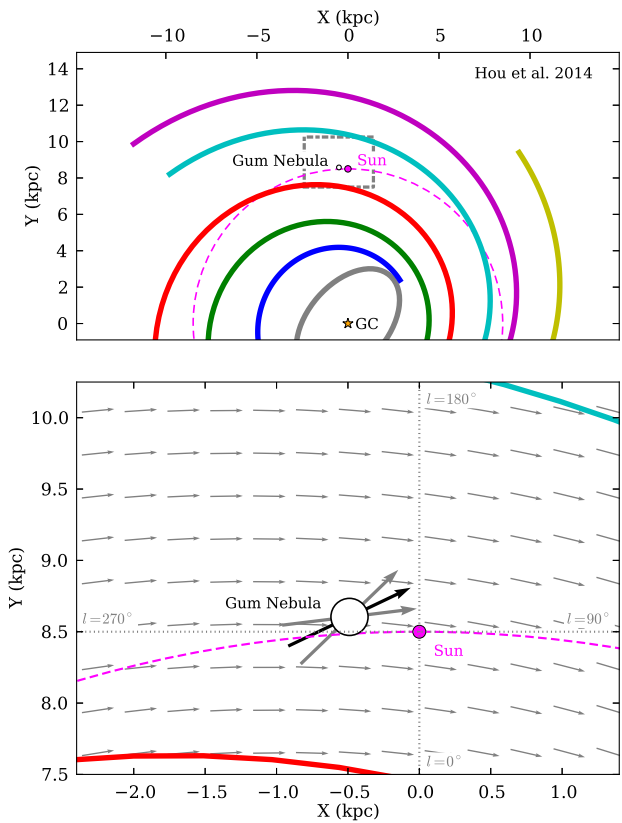}
  \caption{\small {\it Top panel:} Cartoon showing the location of the
    Gum Nebula in the disk of the Milky~Way. The coloured lines
    trace the polynomial-logarithmic spiral arm models of \citet{Hou2014}. 
    A dashed magenta line plots the solar circle at a Galactic radius
    of 8.5\,kpc. {\it Bottom panel:} Illustration of the fitted
    magnetic field orientation around the Gum  Nebula. The
    plot covers the region inside the grey box in the upper panel and
    the Gum nebula is shown by a black circle (to scale). The three
    thick arrows piercing the Gum Nebula show the median (black) and
    $\pm$1-sigma ranges (grey) of field orientation found by fitting
    the RMs in $\S\ref{sec:results_fit}$. The field of thin grey
    vectors depicts the orientation of a spiral magnetic field 
    with a pitch-angle of $\wp=-7.2\degrees$.} 
  \label{fig:field_orientation}
\end{figure}
When viewed face-on to the Galactic disk, the direction of the large-scale
Galactic magnetic field is characterised by the pitch angle, defined as the
deviation from a circular path around the Galactic centre and given by
$\wp={\rm tan}^{-1}(B_{\rm rad}/B_{\rm az})$, where $B_{\rm radial}$
and $B_{\rm az}$ are the radial and azimuthal components of the
ordered field, respectively. In external spiral galaxies the 
magnetic field lines are observed to closely follow the spiral arm
pattern, but the field strength is often greatest in the inter-arm
region (see the examples of \citealt{Beck2005}, \citealt{Fletcher2004}
and \citealt{Patrikeev2006}). In the Milky Way, the ordered disk field is
directed parallel to the disk with a typical strength 
of $B_0=1.5\,-\,2\,\mu{\rm G}$ \citep{Han2006}. 
The pitch angle of the field in the disk has been estimated by multiple
authors using a variety of techniques and has been found to lie between
$-6\degrees$ and $-11.5\degrees$, 
depending on the method used and the volume of the Galaxy
observed. There is also some evidence that $\wp$ may have a radial
dependence, decreasing to almost zero at galactocentric radii greater
than the solar orbit \citep{vanEck2011,Jansson2012}. 
Table~\ref{tab:pitch_angle} summarises the results of individual
studies in the literature.

The pitch angle $\wp$, Galactic longitude $l$ and fitted magnetic
field angle $\Theta$ are related by simple geometry via
$l-180\degrees=\Theta+\wp$. At the Galactic longitude of the Gum
nebula ($l\approx258.0\degrees$) the median pitch angle of
$\wp\approx-7.2\degrees$ from Table~\ref{tab:pitch_angle} implies an
ordered field pointing almost directly towards the observer
($\Theta=85.2\degrees$). Our best-fitting shell models presented in
$\S\ref{sec:results_fit}$ return field directions between
$+43\degrees\lesssim\Theta\lesssim+55\degrees$, equivalent to a
pitch angle range $+23\degrees\lesssim\wp\lesssim+35\degrees$,
substantially different to previous results from the
literature. Taking the $\pm1\sigma$ limits for all models, the local
pitch angle is constrained by our data to $+7\degrees\lesssim
p\lesssim+44\degrees$. This range represents the pitch
angle of the uniform ambient field local to the Gum Nebula. Our
results are illustrated in Fig.~\ref{fig:field_orientation}, which
shows the position of the Gum Nebula with respect to the Galactic
spiral arms \citep{Hou2014} and an ideal uniform field with $\wp=-7.2\degrees$. All of
the investigations referenced in Table~\ref{tab:pitch_angle} calculated
the `global' pitch angle averaged along the line-of-sight to pulsars,
radio-galaxies or stars. All but one study \citep{Pavel2012} covered a
broad swathe in Galactic longitude. As such, the derived pitch angle
is an average over a large fraction of the Galactic disk. By contrast,
the method presented here probes only the magnetic field around the
Gum Nebula, on scales of $\sim260$\,pc (the diameter of the
nebula). 

Only a handful of other studies have used bubbles as probes of
Galactic magnetic field structure: \citet{Kothes2009} studied two SNR
in the Canadian Galactic Plane Survey \citep{Taylor2003} and found
support for an azimuthal disk field, while  \citet{Ransom2010}
performed a similar study using old planetary nebulae, but derived
only the line-of-sight field strength. Most recently
\citet{Whiting2009} and \citet{Savage2013} derived the angle of the
magnetic field by modelling $\hii$~region shells and found field
directions compatible with the mean Galactic field.

We caution that some of the deviation in $\wp$ may
be due to systematic errors in the data. As discussed in
$\S\ref{sec:model_limitations}$, $\Theta$ is sensitive to longitudinal
gradients in RM introduced by contaminating magneto-ionic
material, or by the Galaxy in the background. We have taken all
reasonable steps to identify and eliminate such contamination,
however, a definitive correction requires much more finely sampled and accurate
grid of RMs. 

\begin{table*}
\caption{Studies of ordered magnetic field pitch angle in the literature.}
\label{tab:pitch_angle}
\begin{tabular}{clll}
  \tableline
  \tableline
  Pitch Angle & Reference & Notes \\
  \tableline
  $+16\degrees\pm4\degrees$ & \citet{Inoue1981} & Radio-galaxies, $<2$\,kpc, Orion arm \\
  $-6\degrees$ & \citet{Vallee1988} & Pulsars, few kpc, Sagittarius
  and Perseus arms\\
  $-8.2\degrees\pm0.5\degrees$ & \citet{Han1994} & Pulsars (thin disk)
  and radio-galaxies (thick disk) $\sim3$\,kpc\\
  $-8\degrees$ & \citet{Han1999} & Pulsar RMs, $\sim15$\,kpc\\
  $-7.2\degrees\pm4.1\degrees$ & \citet{Heiles1996} & Starlight
  polarisation, few kpc\\
  $-11.5\degrees$ & \citet{vanEck2011} & Radio-galaxies, Galactic
  sector average\\
  $-6\degrees\pm2\degrees$ & \citet{Pavel2012} & Radio-galaxies,
  average along $l=150\degrees$ \\
  \tableline
\end{tabular}
\end{table*}

The best fitting ambient magnetic field strength of
$B_0=3.9^{+4.9}_{-2.2}\,\muG$ is within the range of $2\,\muG$ to 
$15\,\muG$ observed by similar studies of $\hii$~regions
\citep{Gaensler2001,Harvey-Smith2011}. As shown in
Fig.~\ref{fig:triangle_Sun}, $B_0$ is correlated with $f$, 
which is very poorly constrained by the data. The value of $B_0$
above is reported for $f=0.3$, the mean of the marginalised likelihood
distribution. If instead $f$ is set to the most likely value of
$f=0.24$ then $B_0\approx5\,\muG$. The strength of the field within
the shell depends on the position, and varies between the ambient
level and a maximum value of $X\,\times B_0\approx23\,\muG$ when the
$\overrightarrow{B_0}$ lies parallel to the edge of the nebula.

In summary, the pitch angle of the ordered magnetic field threading
the Gum Nebula ($7\degrees\lesssim \wp\lesssim44\degrees$) is
significantly different to previous measurements, most of which were
averaged over kiloparsec-sized volumes. Few small scale measurements
of the field in the diffuse ionised medium exist, so this result may
represent typical deviations on scales of a few hundred
parsecs. Indeed, \citet{Frisch2012} measured even larger deviations in
the ordered magnetic field in the vicinity of the Sun ($<40$\,pc),
consistent with a scenario where the local ISM is a fragment of the
Loop\,{\rm I} superbubble. Such deviations have also been observed in
external galaxies, for example \citet{Heald2012} detected a
significant RM gradient in the spiral galaxy NGC\,6946,
tracing a irregularity in the vertical component of the ordered magnetic
field. The deviation is directly associated with a hole in the $\hi$
image and may be ubiquitous feature of star-forming
galaxies. Expanding bubbles in the disk may also be responsible for
carrying the small-scale turbulent magnetic 
field into the halo, preventing quenching of the dynamo process
 and allowing the mean magnetic field to saturate at a strength
 comparable to equipartition with the turbulent kinetic
 energy \citep{Shukurov2006}. More accurate and better-sampled RMs
are required to confirm our result and eliminate systematic
uncertainties. The strength of the ambient field around the Gum Nebula
is comparable to average values of $2\,-\,4\,\muG$ measured for the
Galaxy as a whole \citep{Han2006} and within $\hii$~regions.

\subsection{Implications of the fitted compression factor}
The best fitting models presented in $\S\ref{sec:results_fit}$
constrain the jump in density at the edge of the nebula,
assuming the shell is 100\,percent ionised (by stellar radiation in
the case of a $\hii$~region or wind-blown-bubble, or by the
shock-front in the case of a SNR). The
fitted value for the compression factor assuming a flat Galactic
background is $X=1.1\,^{+0.5}_{-0.3}$ and assuming a gradient is
$X=6.0\,^{+5.1}_{-2.5}$. At the very least, 
both values imply that the gas within the shell is only moderately
compressed compared to the ISM external to the nebula.

The current consensus in the literature is that the nebula is an old
supernova remnant (see $\S\ref{sec:gum_history}$). SNR pass through
three distinct evolutionary phases \citep{Woltjer1972} before
dissipating: 1) free 
expansion ($t\lesssim300$\,yr), where the swept-up mass is much less
than the ejected mass and the expansion is dominated by the explosion;
2) the Taylor-Sedov phase ($300\,{\rm yr}\lesssim t\lesssim20000\,{\rm
  yr}$), where the swept-up mass dominates and the blast wave expands
adiabatically and 3) the snow-plough phase ($20000\,{\rm yr}\lesssim
t\lesssim1\,{\rm Myr}$), when thermal cooling has become effective and
the shock front decelerates, sweeping up a dense 
shell. According to the widely used models of \citet{Chevalier1974}
the $\sim260$\,pc diameter of the Gum Nebula implies an age of
$\sim1$\,Myr, which would be old indeed for an SNR. At this late stage
of evolution the shock front is expected to cool radiatively, leading
to a compression factor much greater than the values derived here
\citep{Cioffi1988,Cox1999,Reynolds2011}. Efficient cosmic-ray
acceleration processes may also act to increase the compression factor
\citep{Vink2012}. At times $t\gtrsim1$\,Myr, SNR expansion is expected
to slow down to the ambient sound speed (typically $\sim10\,\kms$) and
merge with the ISM, although the exact details of this process are not
clear \citep{Pittard2003}. The expansion velocity measured from
optical spectroscopy towards the Gum Nebula is $v\lesssim10\,\kms$
(\citealt{Srinivasan1987}, \citealt{Sahu1993}). This slow speed and
moderately low compression factor, combined with the flat $n_e$
profile derived in $\S\ref{sec:ne}$ make it unlikely that the nebula
seen in $\halpha$  stems solely from a supernova origin. Instead, the
detected $\halpha$ emission and RM-signature are consistent with an
ionisation front moving at subsonic speeds into the ISM. Both a
classical $\hii$~region and wind-blown-bubble are bounded by an
ionisation front, so we consider these models in turn below. We cannot
completely rule out the old SNR origin as during the dissipation
stage the shell likely re-expands, leading to a decrease in density
and hence compression factor. However, at this stage, we would also
expect the shell to loose cohesion as it merges with the ISM and this
is not seen in the $\halpha$ data towards the Gum Nebula.

The fundamental theory of expanding, over-pressurised $\hii$~regions
was set out by \citet{Stromgren1939}, \citet{Kahn1954} and
\citet{Oort1954}. After the initial formation phase the $\hii$~region
expands approximately isothermally \citep{Dyson1995} until pressure
equilibrium is reached. For a high-mass O-type star this does not happen
within its stellar lifetime. The D-type ionisation front moves into the
ISM at the local sound speed $\lesssim10\,\kms$ and has a three part
structure, consisting of a thin layer of shocked neutral gas separating
the unshocked neutral gas from the ionised gas within the
$\hii$~region. This classical description produces a spherical ionised
region of approximately constant electron-density, at odds with the
observed shell-like structure of the Gum Nebula seen in $\halpha$
emission. One possible way to reconcile the model with the data is if
the ionisation front is expanding into the walls of a pre-existing
cavity. This explanation was first proposed by  \citet{Reynolds1976a}
whereby the ultraviolet flux from the central stars is ionising the
walls of a void and illuminating the ambient magnetic field in the ISM
local to the Gum region. Such a cavity could have been formed by an
older supernova explosion, or evacuated by an older generation of stars.

The ionised cavity explanation is somewhat contrived, especially since
the wind-blown-bubble model can naturally account for the structure
and physical parameters of the Gum Nebula measured from observations
to date. The star $\zeta$\,Puppis is known to drive a powerful stellar
wind, as is $\gamma^2$\,Velorum and the Vela OB2-association (which
may also lie inside the Gum Nebula). The \citet{Weaver1977}
description of a bubble blown by a high-mass star predicts a
shell-like $\hii$~region surrounding a region of shocked stellar
wind. Indeed, \citet{Weaver1977} offer the Gum Nebula as a prototype
wind-blown-bubble powered by the strong stellar winds from
$\zeta$\,Puppis. Figure~3 in their paper illustrates the temperature 
and density profile of a typical bubble. The density jump across the
outer boundary of the $\hii$ shell is $X\approx1.5$, broadly
consistent with our results. In addition, the density in the interior
is low at $n_e\approx0.05\,\cmmthree$ (as is known to be the case for
the Gum Nebula) and in the shell is $n_e\approx2.5\,\cmmthree$,
comparable to measurements in this work. Equation~69 in
\citet{Weaver1977} describes the density in the shell compared to the
ambient upstream density
\begin{equation}
  n_e = n_0\,(V_2^2 + C_0^2)\,/\,C_s^2,
\end{equation}
where $V_2$ is the shell expansion velocity, $C_0$ is the speed of
sound in the ISM and $C_s$ is the sound speed in the shell. For
typical values used in the \citet{Weaver1977} model of the Gum Nebula
($V_2\approx8\,\kms$, $C_0\approx1\,\kms$, $C_s\approx10\,\kms$) then
$X=n_e/n_0=6.5$ in line with our best fitting shell model assuming a
background RM-gradient (see $\S\ref{sec:RMs_selected}$).

In summary, we believe our results point to the wind-blown-bubble
model as the most likely explanation for the origin of the Gum
Nebula.

\begin{figure*}
  \centering
  \includegraphics[width=17.0cm, angle=0, trim=0 0 0 0]{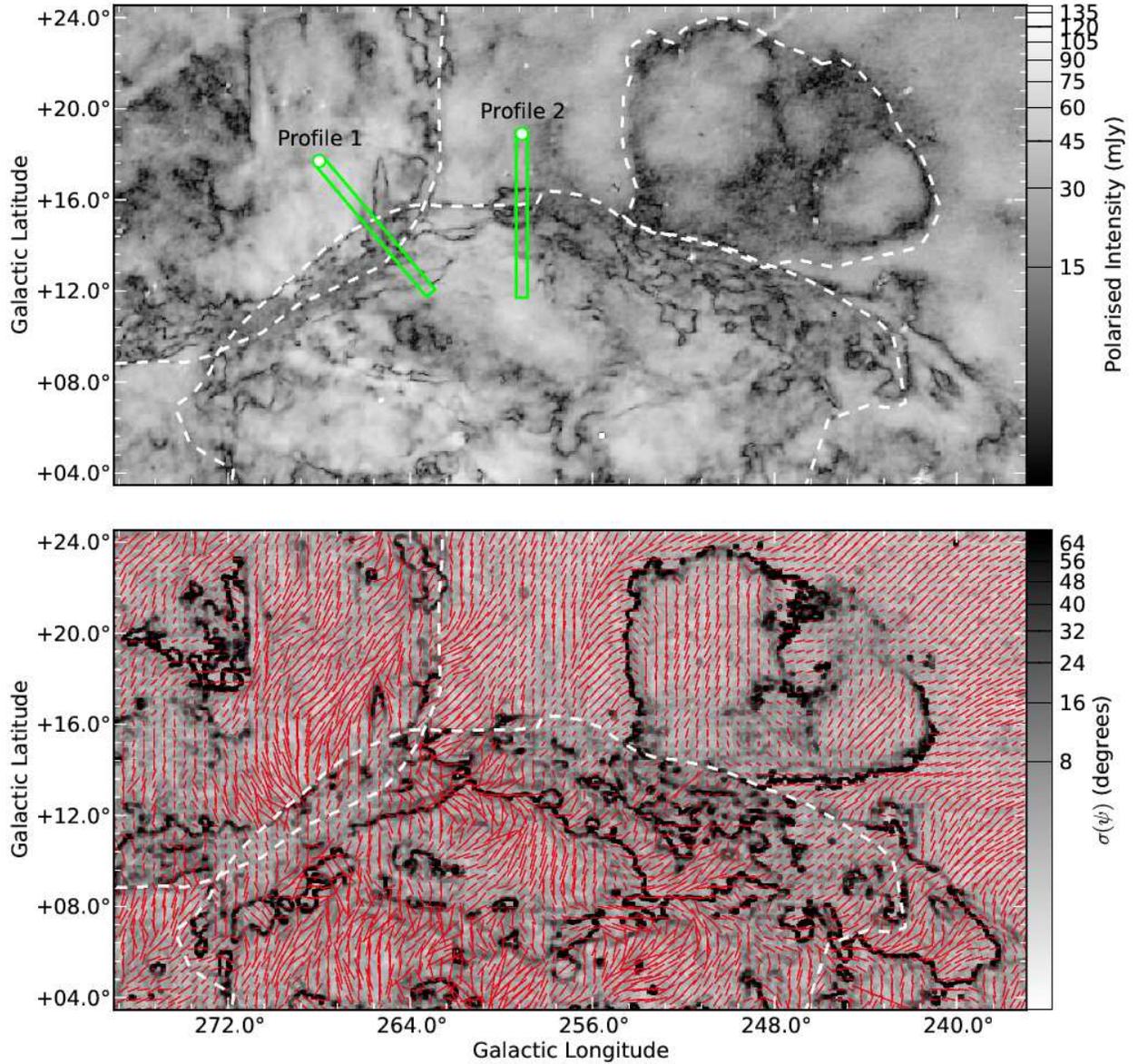}
  \caption{\small {\it Top panel:} Map of the 2.3\,GHz polarised
    intensity for the upper Gum Nebula. Green rectangles show
    where profiles have been extracted, in Fig.~\ref{fig:PI_profiles}
    and dotted lines outline the Gum Nebula and Antlia SNR. The
    profiles run from $(l,b)=(268.0\degrees, 17.7\degrees)$ to ,
    $(263.1\degrees, 11.9\degrees)$ and $(259.1\degrees,
    18.9\degrees)$ to $(259.1\degrees, 11.7\degrees)$ for profiles 1
    and 2, respectively. {\it Bottom panel:} Map of the dispersion in
    polarisation angle $\sigma(\psi)$. The value of $\sigma(\psi)$ in
    each pixel has been 
    calculated from the standard deviation within a beam-sized
    aperture. Dark canals correspond to regions where the polarisation
    angle varies by close to $90\degrees$ within a beam. The red bars
    over-plotted on the image every 7th pixel illustrate the
    orientation of the polarisation vectors (magnetic vector, not
    corrected for RM).}  
  \label{fig:PI_PA_map}
\end{figure*}

\subsection{Pressure and evolutionary state}
The ratio $\beta_{\rm th}=P_{\rm th}/P_{\rm mag}$ of thermal to
magnetic pressures in an $\hii$~region can indicate whether the object
is still evolving or has reached an equilibrium state. In a young
$\hii$ region, the thermal pressure dominates and drives the expansion
of the warm ionised gas into the ISM, sweeping up ambient gas before
it. If the region is threaded by a uniform magnetic field,
flux-freezing in the ionised gas will resist expansion perpendicular
to the field lines. Over time, as the $\hii$~region expands, the
thermal pressure decreases and the magnetic pressure increases, so
ratios closer to unity imply an older object. Magnetic pressure is given by
\begin{equation}
   P_{\rm mag} = B_0^2/\,(8\,\pi)~~~{\rm dyne\,cm^{-2}},
\end{equation} 
where $B_0$ is the total magnetic field strength in $\muG$. The thermal
pressure in an ionised gas at temperature ${\rm T_e}$ (in~K) and
density $n_e$ (in~$\cmmthree$) is
\begin{equation}
  P_{\rm th} = 2\,n_e\,k\,{\rm T_e}~~~{\rm dyne\,cm^{-2}},
\end{equation}
where $k$ is  Boltzmann's constant. Assuming ${\rm T_e}=8000$\,K
and $n_e=1.4\,\cmmthree$ for the Gum Nebula, we find
$P_{\rm th}=2.9\times10^{-12}\,{\rm dyne\,cm^{-2}}$ compared to
$P_{\rm mag}=6.1\times10^{-13}\,{\rm dyne\,cm^{-2}}$. The ratio of thermal
to magnetic pressure $\beta_{\rm th}=P_{\rm th}/P_{\rm mag}=4.8$
suggests that the dynamics of the shell are dominated by thermal
motions, i.e., the magnetic field is too weak to shape the overall
morphology of the ionised gas. This result is in keeping with the
magneto-hydrodynamic (MHD) simulations of \citet{Krumholz2007} and
\citet{Arthur2011}, who find that the thermal pressure of the ionised
gas shapes the evolution on time-scales of several Myr. Both
simulations calculate similar field-strengths and gas-densities to
what we derive for the Gum Nebula. Our values for $n_e$, $B_0$ and
$\beta_{\rm th}$ also sit in the middle of the range found by
\citet{Harvey-Smith2011} in their survey of high-latitude evolved
$\hii$~regions.

\begin{figure}
  \centering
  \includegraphics[width=7.5cm, angle=0, trim=0 0 0 0]{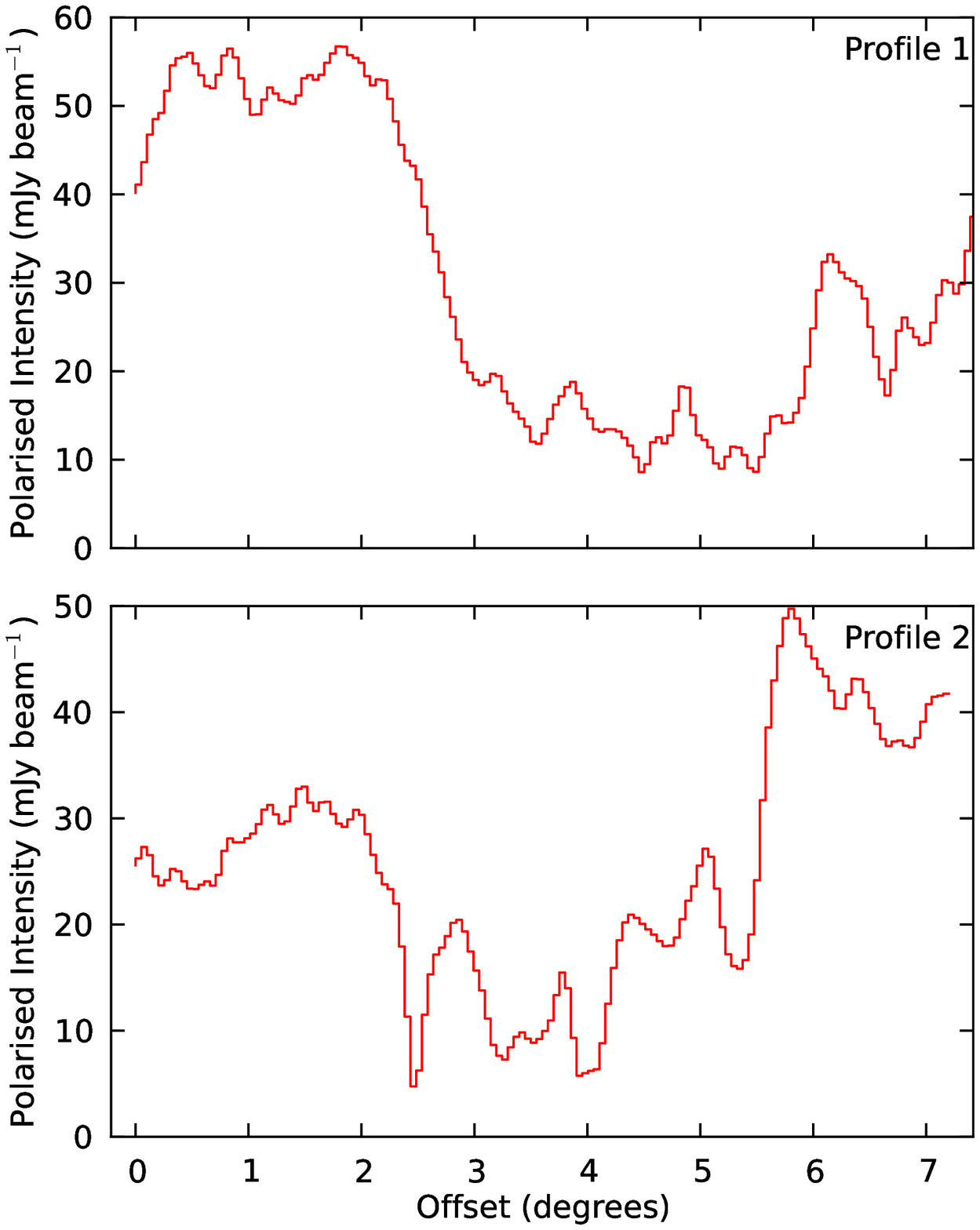}
  \caption{\small Two profiles extracted from the map of the 2.3\,GHz
    polarised intensity. The tracks along which the profiles have
    been extracted are illustrated by the green boxes in
    Fig.~\ref{fig:PI_PA_map} and the starting positions marked by
    circles. } 
  \label{fig:PI_profiles}
\end{figure}
\begin{figure}
  \centering
  \includegraphics[width=8.0cm, angle=0, trim=0 0 0 0]{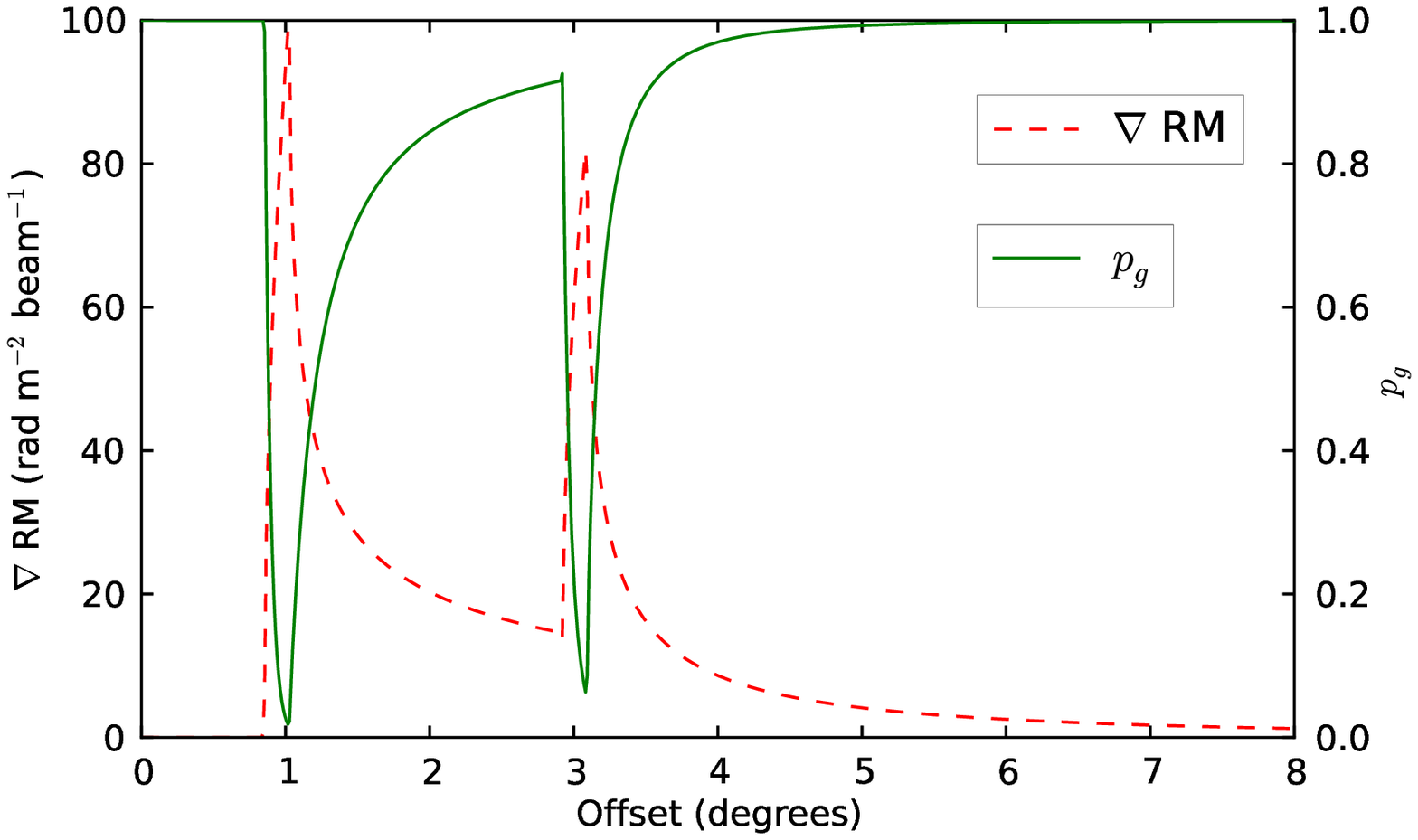}
  \caption{\small The depolarising effect of the spatial
    gradient in RM at the inner and outer edge of an ionised
    shell. The dotted/red line shows the gradient in rotation measure
    ($d{\rm RM}/dr$) in units of $\rmtwo\,{\rm beam}^{-1}$ as a
    function of position, crossing the shell from outside to the
    interior. The solid/green line shows the corresponding degree of
    polarisation calculated from Equation~\ref{eqn:depolGrad}.}
  \label{fig:gradRM_profiles}
\end{figure}

\subsection{Radio Spectral Index}\label{sec:SI}
The Gum Nebula is a prominent foreground feature in the recently
released {\it Planck} all-sky radio-continuum maps
\citep{Planck2013overview}. The $\sim36\degree$ angular diameter shell
is visible in the 28.4\,GHz image at the $\sim10\sigma$ level and in the
44.1\,GHz image at the $\sim3\sigma$ level, implying a flat spectral
index and a significant thermal component to the emission. A more
quantative method of determining the spectral index of diffuse
emission is provided by the temperature-temperature (TT) plot (e.g.,
\citealt{Tian2006}). An x-y plot of the flux densities within a
sampling aperture results in a scatter plot and the slope of a
straight line fit to the data gives the spectral index
$S=S_0\nu^{\alpha}$. The main advantage of a TT-plot is that large
scale offsets in the background emission are automatically compensated
for, assuming that the background does not vary significantly within
the sampling box. We find that the spectral index of the Gum Nebula
shell at the brightest region is $\alpha=0.2\pm0.2$, based on a
comparison of the 28.4\,GHz {\it Planck} data and 2.3\,GHz S-PASS
data; consistent with thermal free-free radio emission. We note that
parts of the Gum Nebula are also faintly visible in the 408\,MHz
radio-continuum map of \citet{Haslam1982}, implying a mixture of
thermal and synchrotron emission in places. 

Further investigation of the spectral index throughout the Gum Nebula
is beyond the scope of this work and should be the subject of a
separate paper. There now exist many wide-angle maps of
radio-continuum emission covering the Gum Nebula, including historical
data (e.g., 45\,MHz: \citealt{Maeda1999}, 1.4\,GHz: \citealt{Reich2001}
and \citealt{Calabretta2014}, 300\,MHz\,-\,1.8\,GHz:
\citealt{Wolleben2009}, 5\,GHz: \citealt{King2010}, 23\,-\,94\,GHz:
\citealt{Bennett2013}) and new diffuse polarisation maps from
{\it Planck} and the Murchison Widefield Array
\citep{Tingay2013}. Future investigations combining these datasets
will be capable of disentangling emission due to synchrotron,
free-free and `spinning dust' processes across the region.

\subsection{Polarised 2.3\,GHz radio-continuum emission}
The properties of the diffuse polarised 2.3\,GHz radio emission
provide complementary information to the RMs of background
radio-galaxies. In particular, analysis of the polarised intensity
and the angle of the linear polarisation vector can yield
information on the geometry and the level of turbulence in the ionised
gas. 

Fig.~\ref{fig:PI_PA_map} presents two views of the polarised
2.3\,GHz emission centred on the upper shell of the Gum Nebula. The
top panel displays a high resolution image of the polarised intensity
($P=\sqrt{U^2 + Q^2}$). The rim of the Gum Nebula stands out as a broad
($\sim2\degrees$ wide) band of depolarisation across the centre of the
image. In addition to the Gum Nebula, two other objects have been
identified in the field. The edge of the Antlia SNR
{$\S\ref{sec:RMs_selected}$, \citep{McCullough2002} is visible in the
  upper-left quadrant as an arc of weaker depolarisation and narrow
  canal-like features.  Such canals 
trace regions where the polarisation angle $\psi$ varies significantly
across a telescope beam, leading to depolarisation in the
receiver \citep{Fletcher2006}. The Antlia SNR overlaps the Gum Nebula
between $163\degrees<l<172\degrees$, where the $\halpha$ emission is
brightest (see Fig.~\ref{fig:gum_summary_halpha}). Also visible are
prominent depolarisation canals from a pair of shells in the
upper-right quadrant of the image (see
$\S\ref{sec:RMs_selected}$). \citet{Iacobelli2014} have analysed the
spatial polarisation gradient \citep{Gaensler2011} in the S-PASS data and have
identified these features as the signature of weak shocks
(see~\citealt{Burkhart2012}). It is not known if either object is
physically interacting with the Gum Nebula, or is simply seen in
projection along the line-of-sight. The bottom panel plots the
polarisation angle of the electric vector (red bars) over an image of
the spatial dispersion in the polarisation angle (\,$\sigma(\psi)={\rm
  stdev}(\psi)$, \citealt{Hildebrand2009}). Maps of $\sigma(\psi)$
have been shown to highlight depolarisation canals
\citep{Planck2013polover} and are useful way of visualising where the
polarisation vectors are homogeneous or heterogeneous on the sky.
Within the Gum Nebula and Antlia SNR the polarised intensity is
patchy and the polarisation angles are chaotic in comparison to the
slowly varying distribution of angles outside their borders. We
interpret this as a `scrambling' of the smooth synchrotron emission
from the Galaxy in the background by Faraday screens associated with
each object (e.g., \citealt{Carretti2013a}). It is clear from previous
studies (e.g., \citealt{Duncan1996}) and radio-continuum data (see
$\S\ref{sec:SI}$) that emission from the Gum Nebula is dominated by
thermal processes. The shell of the nebula does not emit significant
amounts of synchrotron radiation and so can be analysed as a pure Faraday
screen. Assuming a smooth synchrotron background of polarised synchrotron
radiation from the Galaxy, we can quantify the effects of the screen
by examining how the level of polarisation changes across the edge of
the Gum Nebula.

Fig.~\ref{fig:PI_profiles}
presents two polarised profiles extracted from the polarised intensity
map. Profile\,1 cuts across the brightest region of shell, where it
overlaps the weaker Antlia SNR and Profile\,2 has been extracted from
the upper part of the shell. Profile\,1 drops from a high of $\sim55\,\mjbm$
outside the shell to a low of $15\,\mjbm$ inside the depolarised
region. The typical root-mean-squared intensity in the $P$ image is
$2.2\,{\rm m}\jybm$ and the broad depolarisation band exhibits
polarised emission at a $\sim5\sigma$ level. Assuming the outer value
represents the intrinsic polarised intensity $P_i$, then the degree of
depolarisation implied by the drop to $P_0=15\,\mjbm$ is
$p=P_o/P_i\approx0.27$. The drop in intensity and degree of
depolarisation is similar for Profile\,2.

\subsubsection{Depolarisation}
The causes of depolarisation have been described in detail by \citet{Burn1966},
\citet{Tribble1991} and \citet{Sokoloff1998}. The root cause in all
cases is cancellation between polarisation vectors over some averaging
interval in time, space or frequency.

Bandwidth depolarisation occurs when Faraday rotation causes the
polarisation angle to vary across a frequency averaging window
$\Delta\nu$. The degree of depolarisation due to frequency averaging
is
\begin{equation}
p = \left|\frac{\rm sin\,\Delta\psi}{\Delta\psi}\right|,
\end{equation}
where the change in angle across a band centred on $\nu_0$ is given by
$\Delta\psi = -2\,{\rm RM}\,c^2\,\Delta\nu\,/\,\nu_0^3$.
In the Gum Nebula the maximum RM detected is $\sim350\,\rmtwo$ so the
expected angle change over the 244\,MHz bandwidth is
$\Delta\psi=72\degrees$ and the resultant depolarisation is negligible at
$p=0.75$. A RM of $870\,\rmtwo$ would be necessary to completely depolarise
S-PASS data.

The most likely depolarisation mechanism affecting the S-PASS
data is beam depolarisation. This is caused by variations in $B_{||}$
or $n_e$ on scales much smaller than the beam,
scattering the polarisation angles on adjacent lines of
sight. \citet{Burn1966} quantified this effect in the simplest case of
a uniform slab and found 
\begin{equation}\label{eqn:burn_depol}
  p = {\rm exp}\,(-2\,\sigma_{\rm RM}^2\,\lambda^4),
\end{equation}
where $\sigma_{\rm RM}$ is the RM scatter within a beam after
measurement errors have been accounted 
for. If small-scale random fluctuations are solely responsible for the
observed depolarisation ($p=0.27$) then Equation~\ref{eqn:burn_depol}
predicts an excess scatter of $\sigma_{\rm RM}=47\,\rmtwo$. In
$\S\ref{sec:results_fit}$ we found that the best-fitting model implied
an additional scatter of $\sigma_{\rm RM}=78.6\,\rmtwo$ (called
$\delta({\rm RM})$ in Table~\ref{tab:results}). This fitted value
is an upper-limit on $\sigma_{\rm RM}$ as the RM sampling grid
is very coarse at ${\rm \sim1/degree^2}$, compared to the beam FWHM
of $\Theta_{\rm beam}=10.75'$. We can conclude that the data
is at least consistent with a large fraction of the depolarisation
being due to random fluctuations in $B_{||}$ or $n_e$.

While the average drop in $P$ can be explained (at least in part) by
fluctuations within an ionised Faraday screen, the shell of the
nebula also contains depolarisation canals. These are typically one beam
in width, close to 100 percent depolarised and tend to be aligned
parallel to the edge of the nebula. First discovered by
\citet{Haverkorn2000}, several authors in the last decade
have studied origin of such canals and explored their use as a
diagnostic tool (e.g., \citealt{Fletcher2006}, \citealt{Gaensler2011}
and~\citealt{Burkhart2012}). In particular, \citet{Gaensler2011}
calculated the spatial gradient of the complex Stokes vector
$\overrightarrow{P}=(\overrightarrow{Q},\,\overrightarrow{U})$, whose magnitude $|\nabla\,P|$ describes
the rate at which the polarisation vector traces out a path in the
$Q$\,--\,$U$ plane when moving along a spatial track at a constant
rate. $|\nabla\,P|$ is invariant under arbitrary rotations or
translations (unlike $P$ or $\psi$) and images of $|\nabla\,P|$
reveal a network of filaments in the ionised gas (see
\citealt{Iacobelli2014} for the $|\nabla\,P|$ of the S-PASS data). In
a pure Faraday screen these filaments have  
been shown to trace spatial cusps or jumps in $n_e$ or $B_{||}$, most
likely caused by shock-fronts or turbulent motions in the gas
\citep{Burkhart2012}. Depolarisation canals like those in
Figure~\ref{fig:PI_PA_map} are a subset of filaments that cross the
origin in the $Q$\,--\,$U$ plane. The greatest concentration of canals
occur within the rim of the nebula, lending weight to our conclusion
that turbulent fluctuations in $n_e$ or $B_{||}$ are responsible for
the depolarisation. 

The most prominent canals run along the inner and
outer edges of the depolarised rim and can be explained by the
intrinsic RM-gradient at the edges of the ionised shell. The
amount of depolarisation produced by an RM-gradient is given by
\citet{Sokoloff1998} as 
\begin{equation}\label{eqn:depolGrad}
  p_g = exp\,\left[ - \frac{1}{ln\,2}\, \left(\frac{d{\rm RM}}{dr} \right)^2\,\lambda^4\,\right],
\end{equation}
assuming a Gaussian beam which resolves the
gradient. Fig.~\ref{fig:gradRM_profiles} plots the RM-gradient and
the depolarisation factor calculated from the
Equation~\ref{eqn:depolGrad} for a profile crossing towards the
interior of the ionised shell. From the plot we see that
depolarisation only becomes significant ($p<0.6$) close to peaks in
the gradient. The equation breaks down for resolved gradients, however,
it is clear that narrow depolarisation canals are predicted at the
leading and inner edges of the shell.

In conclusion, the polarisation and depolarisation properties of the
2.3\,GHz S-PASS data are in keeping with the simple ionised shell
model put forward in $\S\ref{sec:mag_bubble}$ and support our
assertion that the Gum Nebula is acting as a Faraday screen. 

\subsection{Comparison to previous studies }
The first dedicated magnetic field measurements of the Gum Nebula
were obtained by \citet{Vallee1983} via linear polarisation
observations of 35 background extragalactic radio sources. Prior to
that work, large-scale rotation measure excesses in the area were
attributed to a tangential view of the local Orion-spur
spiral arm \citep{Simard1980}. \citet{Vallee1983} claimed that the
distribution of RMs on the sky were not consistent with the arm model
but were a good match for the old SNR model first presented
by \citet{Reynolds1976b}. Their derived line-of-sight magnetic field
strength of $\sim1.3\,\mu$G suggested that a `snow-plough' effect
alone was responsible for sweeping up gas, and hence the magnetic
field lines. In later work, \citet{Duncan1996} cast doubt on the
significance of the Vallee model, pointing out that the statistical
uncertainty in the data used therein was comparable to the mean RM
value.  The model presented in this paper is broadly consistent with the
\citet{Vallee1983} result, but is considerably more sophisticated and
includes much better sampled measurements of RM and $n_e$. We also
derive independent values for the shell thickness and compression
factor, which \citet{Vallee1983} did not provide.

Magnetic field strengths in ionised bubbles have been measured by a
number of recent studies in the literature. \citet{Whiting2009} and
\citet{Savage2013} used a similar technique to the one presented here
to study the bubble surrounding the Cygnus OB1 association and the
Rosette nebula, respectively. \citet{Whiting2009} suggested that the
observed Faraday `anomaly' was caused by a wind-blown bubble, but with
only nine RMs 
they could not confirm the compression factor predicted by the strong
shock. On the other hand, \citet{Savage2013} modelled RMs seen through
the Rosette nebula as a limb-brightened ionised shell and obtained a
considerably better fit when fixing $X=4$ compared to $X=1$. Recently,
\citet{Harvey-Smith2011} studied the line of sight magnetic fields in
five large-diameter $\hii$~regions offset from the Galactic
plane. They derived field strengths from $\sim3\,-\,11\,\mu$G, but
found no evidence of compression at the edges of these relatively
evolved $\hii$~regions. 

\section{Summary and Conclusions}\label{sec:summary}
We have developed a simple model of the Gum Nebula as an expanding
ionised shell threaded by a uniform magnetic field. Drawing upon
the RM catalogue of \citet{Taylor2009} and the $\halpha$ image of
\citet{Finkbeiner2003}, we used a maximum-likelihood MCMC analysis to
derive the magneto-ionic shell parameters in the upper hemisphere of
the nebula. We compared the best-fitting models to polarised
2.3\,GHz radio-continuum emission from the S-PASS project. Our
conclusions are as follows:
\begin{enumerate}
  \item The RM and EM data covering the upper hemisphere of the Gum Nebula
    ($b>5\degrees$) are well-fitted by a simple ionised shell. Assuming a
    large-scale RM-background from the \citet{Sun2008} model of the
    Galaxy, the best-fitting shell has an angular radius
    $\phi_{\rm outer}=22.7\degrees^{+0.1}_{-0.1}$, shell thickness 
    $dr=18.5^{+1.5}_{-1.4}\,{\rm pc}$, ambient magnetic field strength
    $B_0=3.9^{+4.9}_{-2.2}\,\muG$, electron density $n_e=1.4^{+0.4}_{-0.4}\,{\rm
      cm^{-3}}$ and filling factor $f=0.3^{+0.3}_{-0.1}$ 
  \item We constrain the pitch angle of the uniform magnetic field to
    values over the range $+7\degrees\lesssim\wp\lesssim+44\degrees$,
    significantly different from previously derived values
    ($\wp\approx-7\degrees$) averaged over much larger volumes of the
    Galactic disk (scales of several kpc versus $\sim260$\,pc for the
    Gum Nebula). Our fitted values are sensitive to contamination of
    the RMs by intervening magneto-ionic objects, however, we have
    corrected the catalogues to the full extent allowed by the
    available data. This represents one of the few measurements of
    local magnetic field orientation in the Milky Way.
  \item We find that the compression factor $X=n_e/n_0$ at the edge of
    the $\halpha$ shell is $X=6.0\,^{+5.1}_{-2.5}$, assuming an
    RM-background from \citet{Sun2008}. This value is much lower than
    expected if the Gum Nebula were an old SNR cooling radiatively.
    We believe that the most likely explanation for the nebula is a
    wind-blown-bubble driven by a cluster of high-mass stars. The slow
    expansion velocity ($\lesssim10\,\kms$), low excitation conditions
    and lack of radio-synchrotron emission from the rim is consistent
    with our hypothesis. 
  \item The strength of the ordered magnetic field $B_0$ is not well
    measured as it is degenerate with the ill-constrained filling
    factor $f$. We derive a value of $B_0=3.9^{+4.9}_{-2.2}\,\muG$, in
    line with the strength of the ambient Galactic field and also
    comparable with values measured towards $\hii$~regions by previous
    authors.
  \item Viewed in 2.3\,GHz radio-continuum, the upper shell of the Gum
    Nebula exhibits a distinctive band of depolarised emission. We
    find that the dominant depolarising mechanism is likely due to
    fluctuations in $n_e$ and the random component of
    $\overrightarrow{B}_o$ on scales much smaller than the $10.75'$
    beam. The depolarised canal features observed at the boundary of
    the band are consistent with being caused by RM-gradients at the
    edge of the ionised shell.
\end{enumerate}

The study presented here illustrates how even large objects,
well-sampled by RMs, require great care to disentangle from confusing
sources. The next generation of surveys planned
for the Square Kilometer Array and precursors instruments will enable
similar studies of many more Galactic objects. From 2016 onwards the
Australia Square Kilometre Array Pathfinder (ASKAP,
\citealt{Johnston2007}) POSSUM (Polarisation Sky Survey of the
Universe's Magnetism) project \citep{Gaensler2010} will survey the
southern sky at 1\,GHz and deliver a RM-grid with $\sim$100 polarised
sources per square degree ($\sim100$ times the source density of the
NVSS). Once the new catalogue becomes available it will be possible to
identify and correct for smaller Faraday-active objects with greater
accuracy and confidence. This work serves as a rehersal for future
studies and highlights the challenges involved.


\acknowledgments
We would like to thank the anonymous referee, whose thorough comments
significantly improved this paper. We are very grateful to Roland
Kothes and James Allison for useful discussions on the physics of
bubbles and MCMC analysis, respectively. We thank Tom Landecker for
his careful reading of the manuscript and for his valuable
comments. We are also indebted to Rainer Beck, Marijke Haverkorn,
Wolfgang Reich and Julian Pittard for detailed suggestions. CRP, BMG
and XHS were supported by the Australian Research Council through grant
FL100100114. Parts of this research were conducted by the Australian
Research Council Centre of Excellence for All-sky Astrophysics
(CAASTRO), through project number CE110001020. This work has been
carried out in the framework of the S-band Polarisation All Sky Survey
(S-PASS) collaboration. The Parkes Radio Telescope is part of the
Australia Telescope National Facility, which is funded by the
Commonwealth of Australia for operation as a National Facility managed
by CSIRO. The Southern H-Alpha Sky Survey Atlas (SHASSA) is supported
by the National Science Foundation. This research has made use of
NASA's Astrophysics Data System. This research also made use of
Astropy, a community-developed core Python package for Astronomy
\citep{Astropy2013}.



\bibliographystyle{apj}
\bibliography{gum_Bfield}

\appendix

\section{Geometry of Simple Ionised Shell Model}\label{app:shell_geom}
\begin{figure*}
  \centering
  \includegraphics[width=16.0cm, trim=0 0 0 0]{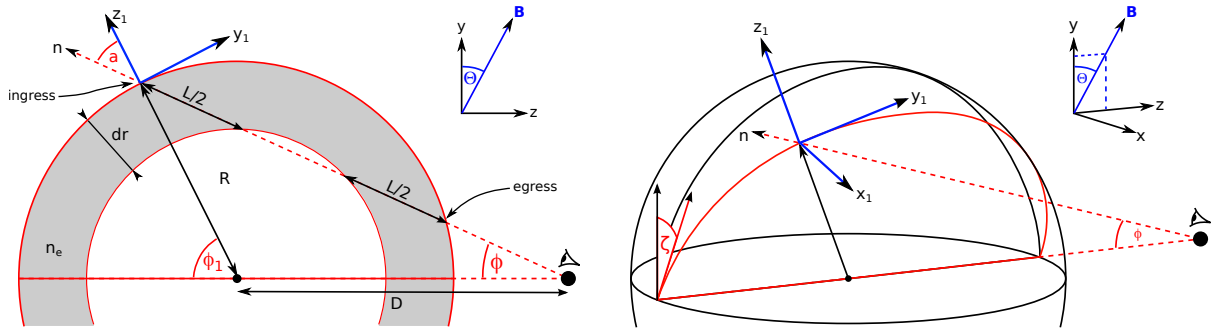}
  \caption{\small Geometry of an ionised shell in the near-field. We
    assume a constant electron density $n_e$ in the shell, which is
    threaded with a uniform magnetic field. The field is confined to the
    y-z plane, representing the Galactic disk, and only the
    sky pitch-angle $\Theta$ may be varied. }
  \label{fig:bubble_geom}
\end{figure*}
The geometry of our simple shell model is shown schematically in
Fig.~\ref{fig:bubble_geom}. It consists of a spherical ionised region
of radius $R$ and 
thickness $dr$, containing thermal electrons at an average density
$n_e$. The edge of the shell subtends an angular radius $\phi_{\rm
  outer}={\rm sin}^{-1}(R/D)$, where $D$ is the distance from
the observer to the geometric centre. The ionised gas is threaded by a
uniform magnetic field $\overrightarrow{B_0}$, whose vector is
confined to the y-z plane representing the Galactic disk. Field
lines make an angle $\Theta$ to the plane of the sky in the direction
of the bubble centre. From the perspective of an external observer,
the angle $\zeta$ describes the orientation of the sight-line to the
y-z plane. The observed RM is given by Equation~\ref{eqn:RM_1} and in
the simplest case ($\zeta=0$, Fig.~\ref{fig:bubble_geom}\,-\,{\it
  left}) depends only on the line-of-sight field strength $B_{||}$
and the path-length L along a chord through the shell:
\begin{equation}
  B_{||} = - B_0\,{\rm sin}(\Theta-\phi),
\end{equation}
\begin{equation}\label{eqn:path_chord}
  L(\phi) = 2\,\sqrt{D^2 \left[{\rm sin}^2(\phi_{\rm outer}) - {\rm sin}^2(\phi)\right]}.
\end{equation}
If the shell is compressing the ionised gas at the
leading edge then the components of $\overrightarrow{B_0}$ tangent to
the surface will be amplified by a factor $X$. To model this effect we
decompose $\overrightarrow{B_0}$ at each point on the surface into 
vector components ($B_{x_1}$, $B_{y_1}$, $B_{z_1}$) along 
the axes $x_1$, $y_1$ and $z_1$ such that
\begin{align}
  \label{eqn:b_surface_sphere1}
  B_{x_1} & = B_0~{\rm cos}(\Theta)\,{\rm sin}(\zeta),\\
  \label{eqn:b_surface_sphere2}
  B_{y_1} & = B_0~\left[~{\rm cos}(\Theta)\,{\rm cos}(\zeta)\,{\rm cos}(\phi_1) +
    {\rm sin}(\Theta)\,{\rm sin}(\phi_1)~\right],\\
  \label{eqn:b_surface_sphere3}
  B_{z_1} & = B_0~\left[~{\rm cos}(\Theta)\,{\rm cos}(\zeta)\,{\rm sin}(\phi_1) -
    {\rm sin}(\Theta)\,{\rm cos}(\phi_1)~\right].
\end{align}
The observed magnetic field strength is then the vector sum along the
line-of-sight at both the ingress ($B_i$) and
egress ($B_e$) points. From the geometry we find 
\begin{align}
  B_i & = X\,B_{y_1}~{\rm sin}(a) - B_{z_1}~{\rm cos}(a),\\
  B_e & = X\,B_{y_1}~{\rm sin}(a) + B_{z_1}~{\rm cos}(a),
\end{align}
where $a = {\rm sin}^{-1}(\phi/\phi_{\rm outer})$. When calculating
$B_i$ the angle $\phi_1$ in
Equations~\ref{eqn:b_surface_sphere1}\,-\,\ref{eqn:b_surface_sphere3}
is sampled over the far side of the sphere according to
$\phi_1=a+\phi$. Similarly, when calculating $B_e$, $\phi_1$ is
sampled over the near side: $\phi_1 = 180\degrees -a +\phi$. 
The final line-of-sight magnetic field strength is then given by
\begin{equation}\label{eqn:blos}
  B_{||} = (B_i + B_e)\,/\,2,
\end{equation}
assuming $B_i$ and $B_e$ are constant along the path connecting
the midpoint of the chord and the surface of the shell. The RM at any
point may then be calculated by combining
Equations~\ref{eqn:path_chord}\,-\,\ref{eqn:blos} with
Equation~\ref{eqn:RM_los}.
\end{document}